\documentclass[transmag]{IEEEtran}

\usepackage{graphicx}
\usepackage{amsmath,bbm,amssymb,amsfonts,amstext,amsopn,sansmath}
\usepackage{cite}
\usepackage{balance}
\usepackage{url}
\usepackage{epsfig}
\usepackage{setspace}
\usepackage{stmaryrd}
\usepackage{psfrag}	
\usepackage{multirow}
\usepackage{float}
\usepackage[process=auto]{pstool}
\usepackage{etoolbox}
\usepackage{algorithm}
\usepackage{algorithmic}
\allowdisplaybreaks
\usepackage[nolist]{acronym}

\usepackage{latexsym}
\usepackage{graphicx}
\usepackage{amsfonts,amssymb,amsmath}

\usepackage{flushend}
\usepackage{lipsum}

\newcommand{\Trans}{^{\mathsf{T}}}
\newcommand{\Herm}{^{\mathsf{H}}}

\newcommand{\Ptx}{P_{\mathrm{tx}}}
\newcommand{\Prd}{P_{\mathrm{rd}}}
\newcommand{\Ptot}{P_{\mathrm{tot}}}

\newcommand{\Pbb}{P_{\mathrm{bb}}}

\newcommand{\Pamp}{P_{\mathrm{amp}}}
\newcommand{\Prfc}{P_{\mathrm{rfc}}}

\newcommand{\Psw}{P_{\mathrm{sw}}}

\newcommand{\F}{\mathbf{F}}
\newcommand{\B}{\mathbf{B}}
\newcommand{\T}{\mathbf{T}}
\newcommand{\D}{\mathbf{D}}
\newcommand{\Hc}{\mathbf{H}}

\newcommand{\Ex}{\mathbbmss{E}}
\newcommand{\Tr}{\mathrm{trace}}
\newcommand{\Diag}{\mathrm{diag}}

\newcommand{\BLUE}{\color[rgb]{0,0,0}}

\newtheorem{lem}{Lemma}
\newtheorem{remk}{Remark}

\newtheorem{prop}{Proposition}
\newtheorem{corol}{Corollary}

\newtoggle{OneColumn}
\toggletrue{OneColumn}

\newtoggle{arXiv}
\togglefalse{arXiv}

\EndPreamble
\begin{document}
\title{Intelligent Surface-Aided  Transmitter Architectures for  \\ Millimeter Wave  Ultra Massive MIMO  Systems}

\author{
	Vahid Jamali, \IEEEmembership{Member, IEEE}, Antonia M. Tulino, \IEEEmembership{Fellow, IEEE},  Georg Fischer, \IEEEmembership{Senior Member, IEEE},  \\ Ralf M\"uller, \IEEEmembership{Senior Member, IEEE},  and Robert Schober, \IEEEmembership{Fellow, IEEE}
	\thanks{This paper was presented in part at IEEE ICC 2019 \cite{jamali2018scalable}.}
	\thanks{V. Jamali, R. M\"uller, and R. Schober	are with the Institute for Digital Communications at Friedrich-Alexander University Erlangen-N\"urnberg (FAU) (e-mail:
		vahid.jamali@fau.de; ralf.r.mueller@fau.de;	robert.schober@fau.de).
	}
	\thanks{A. M. Tulino is with Nokia Bell Labs, Holmdel, NJ 07733 USA, and also with the Universityà degli Studi di Napoli Federico II, 80138 Naples, Italy
		(e-mail: a.tulino@nokia-bell-labs.com; antoniamaria.tulino@unina.it)}
	\thanks{G. Fischer is with the 	Institute for Electronics Engineering at FAU (e-mail: georg.fischer@fau.de).}}

\IEEEtitleabstractindextext{\begin{abstract}
In this paper, we study two novel massive multiple-input multiple-output (MIMO)  transmitter architectures for millimeter wave (mmWave) communications which comprise few active antennas, each equipped with a dedicated radio frequency (RF) chain, that illuminate a nearby large intelligent reflecting/transmitting surface (IRS/ITS). The IRS (ITS) consists of a large number of low-cost and energy-efficient passive antenna elements 
which are able to reflect (transmit) a phase-shifted version of the incident electromagnetic field. Similar to lens array (LA) antennas, IRS/ITS-aided antenna architectures are energy efficient due to the almost lossless over-the-air connection between the active antennas and the intelligent surface. However, unlike for LA antennas, for which the number of active antennas has to linearly grow with the number of passive elements (i.e., the lens aperture) due to the non-reconfigurablility (i.e., non-intelligence) of the lens, for IRS/ITS-aided antennas, the reconfigurablility of the IRS/ITS facilitates scaling up the number of radiating passive elements without increasing the number of costly and bulky active antennas. We show that the constraints that the precoders for IRS/ITS-aided antennas have to meet differ from those of  conventional MIMO architectures. Taking these constraints into account and exploiting the sparsity of mmWave channels, we design two efficient precoders; one based on maximizing the mutual information and one based on approximating the optimal unconstrained fully digital (FD) precoder via the orthogonal matching pursuit algorithm. Furthermore, we develop a power consumption model for IRS/ITS-aided antennas that takes into account the impacts of the IRS/ITS imperfections, namely the spillover loss, taper loss, aperture loss, and phase shifter loss. Moreover, we study the effect that the various system parameters have on the achievable rate and show that a proper positioning of the active antennas with respect to the IRS/ITS leads to a considerable performance improvement. Our simulation results reveal that unlike conventional MIMO architectures, IRS/ITS-aided antennas are both highly energy efficient and fully scalable in terms of the number of transmitting (passive) antennas. Therefore, IRS/ITS-aided antennas are promising candidates for realizing the potential of mmWave \textit{ultra massive} MIMO communications~in~practice.
\end{abstract}

}

\maketitle

\acresetall
\section{Introduction}

\begin{figure*}
	\centering\vspace{-0.05cm}
	\includegraphics[width=0.9\linewidth]{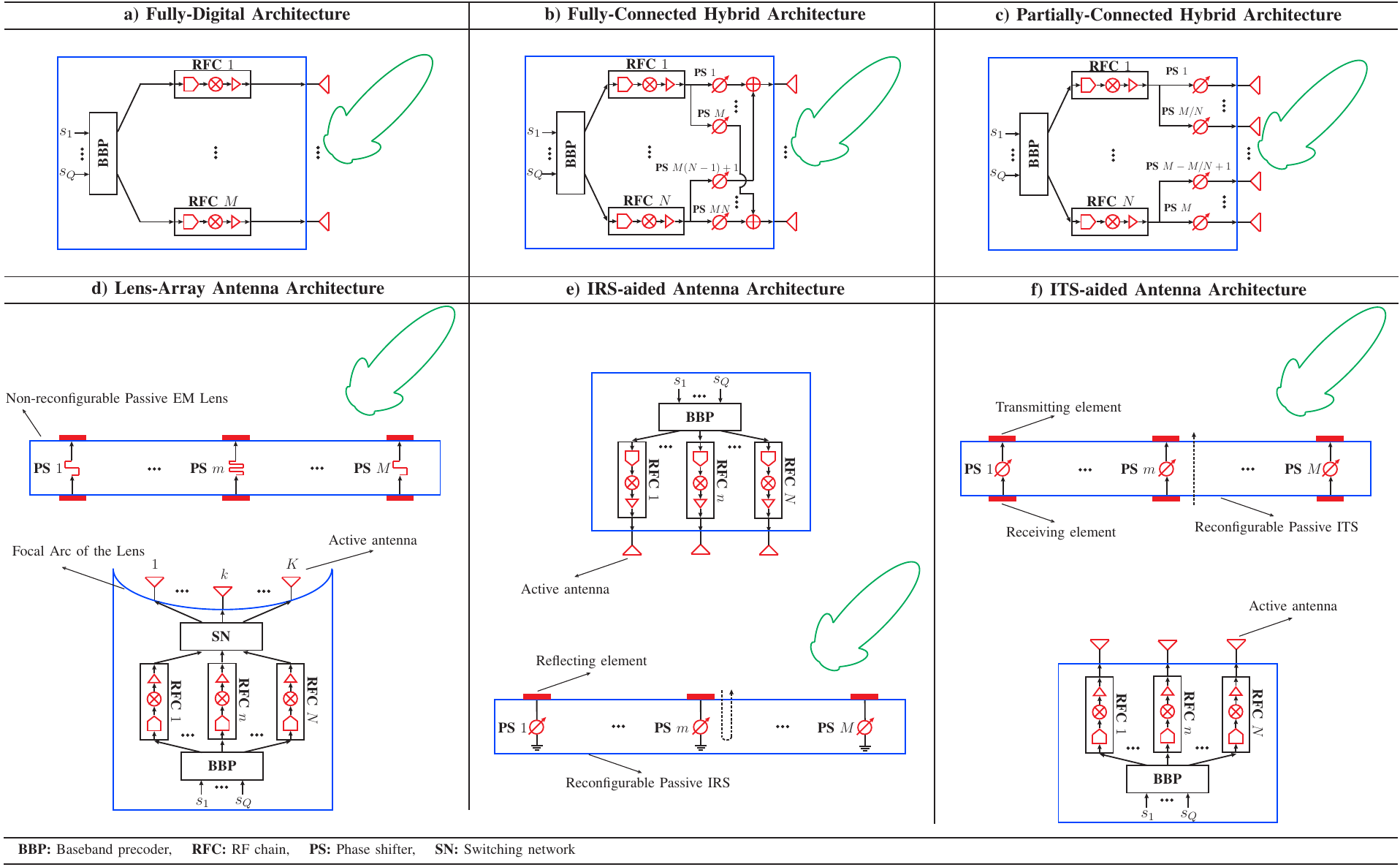} \vspace{-0.02cm}  
	\caption{\BLUE Schematic illustration of the considered massive MIMO architectures. Here, $Q$, $N$, and $M$ denote the numbers of data streams, RF chains, and transmit antennas (passive elements in LA and IRS/ITS-aided antennas), respectively, and $K$ denotes the number of active antennas of the LA architecture. For the LA architecture, the EM lens is a non-reconfigurable surface and consists of fixed phase shifters, whereas the IRS/ITS-aided antenna architectures are equipped with intelligent surfaces employing reconfigurable phase shifters. \vspace{-0.03cm}}
	\label{Fig:SysMod}
\end{figure*}

\IEEEPARstart{M}{illimeter} wave (mmWave) communication systems are promising candidates for realizing  the high data rates expected from the next generation of wireless communication networks \cite{busari2017millimeter,delgado2018feasible}.
These systems will be equipped with a large array of antennas at the transmitter and/or the receiver to cope with the high path loss, limited scattering, and small antenna apertures at mmWave frequencies. However, conventional fully-digital (FD) multiple-input multiple-output (MIMO) systems, which connect each antenna to a dedicated radio frequency (RF) chain, see Fig.~\ref{Fig:SysMod} a), are infeasible for mmWave systems due to the prohibitively high cost and energy consumption of the high resolution analog-to-digital/digital-to-analog converters required for each antenna element \cite{gao2018low}. This has motivated researchers to consider hybrid analog-digital MIMO architectures, which tremendously reduce the number of RF chains by moving some of the signal processing operations into the analog domain \cite{el2014spatially,molisch17,ghanaatian2018feedback}.

Typically, in hybrid MIMO systems, it is assumed that the output of each RF chain is connected to all antennas via an analog network, see Fig.~\ref{Fig:SysMod} b). This architecture is referred to as fully-connected (FC) hybrid MIMO and is able to realize the full beamforming gain of massive antenna arrays. Unfortunately, FC hybrid MIMO is not scalable due to the excessive power consumption of the analog network for large numbers of antennas \cite{yan2018performance}. In particular, the analog network is comprised of RF dividers, combiners, phase shifters, and line connections, which lead to high RF losses and hence reduce energy efficiency. To deal with this issue, partially-connected (PC) hybrid MIMO architectures were proposed in the literature where the output of each RF chain is connected to only a subset of the antennas \cite{gao2018low,gao2016energy}, see Fig.~\ref{Fig:SysMod} c). Thereby, RF combiners are not needed, and the numbers of phase-shifters and RF lines are reduced. Nevertheless, as is shown in \cite{yan2018performance} and also in Section~\ref{Sec:Sim} of this paper, the power consumption of PC hybrid MIMO still scales with the number of antennas in a similar manner as for FC hybrid~MIMO.

To overcome the poor energy efficiency of analog networks, lens array (LA) antennas have been proposed in the literature \cite{zeng2017cost,zeng2018multi,brady2013beamspace,popovic2002multibeam}. LA antennas consist of two main components, namely an electromagnetic (EM) lens and several active antennas, which are connected by an almost lossless wireless link,  see Fig.~\ref{Fig:SysMod} d). EM lenses are phase-shifting devices which can be realized utilizing an array of passive antenna elements \cite{popovic2002multibeam} or continuous aperture phase shifting \cite{brady2013beamspace}. The active antennas are placed on the focal arc of the EM lens and connected to a small number of RF chains via a switching network. The EM lens transmits the signals of different active antennas in different directions \cite{brady2013beamspace,zeng2018multi}. Therefore, due to the sparsity of mmWave channels, only the few active antennas that lead to transmission in the directions of the scatterers in the channel have to be activated. Note that the number of passive antennas (i.e., the effective lens aperture) determines how narrow a beam can be made whereas the number of active antennas limits the number of beam directions (i.e., the resolution of the LA). Hence, as the number of passive elements increases, the number of available active antennas also has to increase in order to maintain a satisfactory performance. Although passive antennas can be small and cheap (e.g., simple patch antennas), active antennas that can transmit with high power are typically bulky and expensive (e.g., horn antennas) \cite{popovic2002multibeam}. Therefore, increasing the number of active antennas constitutes a bottleneck for the scalability of the LA architecture as its implementation becomes costly and bulky for massive MIMO systems.

In order to improve the scalability and energy efficiency of mmWave massive MIMO systems, in this paper, we consider two novel massive MIMO  transmitter architectures\footnote{Following the reciprocity theorem \cite{balanis1982antenna}, the considered transmitter architectures can be also used as receivers. However, for concreteness, in this paper, we focus on transmitter design and leave the receiver design for future work.} which comprise few active antennas and a large intelligent reflecting surface (IRS), see Fig.~\ref{Fig:SysMod} e), or a large intelligent transmitting surface (ITS), Fig.~\ref{Fig:SysMod} f).  
Recently, intelligent surfaces or metasurfaces have been extensively investigated in the literature with the objective to improve the coverage, spectrum, and energy efficiency of wireless communication systems \cite{wu2018intelligent,liaskos2019novel,di2019smart,basar2019wireless,wu2019intelligent,najafi2019intelligent,karasik2020beyond,najafi2020intelligent,jamali2020power,zuo2020intelligent,abdelrahman2017analysis,luyen2019wideband,popovic1998quasi,di2015reconfigurable,pham2019low,hasani2019dual,bereyhi2019papr,bereyhi2020single}. The deployment scenarios of intelligent surfaces can be roughly classified into two categories, namely intelligent surfaces as a part of the wireless channel and as a part of the transceiver architecture\footnote{The term large intelligent surface (LIS) has been also used in the literature  to refer to surfaces that are placed on e.g. walls and comprise a large array of transmitting/receiving antennas, see e.g. \cite{hu2018beyond,jung2020performance}. Unlike the reflecting/transmitting intelligent surfaces considered in this paper, the surfaces in \cite{hu2018beyond,jung2020performance}  neither reflect nor  re-transmit their received signals, instead the input/output of the antenna elements is directly controlled/processed. The main purpose of LISs is to bring the massive MIMO transceivers close to the users and to ensure line-of-sight links.}. For the former case, the waves incident on the intelligent surface are emitted by a transmitter which is located far away, such that the channel between the transmitter and the intelligent surface is subject to fading, cannot be influenced, and has to be estimated for beamforming~design \cite{wu2018intelligent,liaskos2019novel,di2019smart,basar2019wireless,wu2019intelligent,najafi2019intelligent,karasik2020beyond,najafi2020intelligent,zuo2020intelligent}.  In contrast, for the latter case, the intelligent surface is embedded into the transmitter/receiver architecture \cite{abdelrahman2017analysis,luyen2019wideband,popovic1998quasi,di2015reconfigurable,pham2019low,hasani2019dual,bereyhi2019papr,bereyhi2020single}. Therefore, the  waves incident on the intelligent surface are created by physically close active antennas, such that the channel between the active antennas and the intelligent surface  is fixed and can be properly designed during manufacturing.  The focus of this paper is on the second category, namely IRS/ITS-aided transmitter antenna architectures, and their communication-theoretical modeling and system design.

In the considered IRS/ITS-aided MIMO architectures, each active antenna is equipped with a dedicated RF chain and illuminates the IRS/ITS. The IRS (ITS) consists of a large number of low-cost and energy-efficient passive antenna elements which are able to reflect (retransmit) a phase-shifted version of the incident electromagnetic field. In particular, each passive element receives a superposition of the signals transmitted (over the air) by the active antennas and adds a desired phase shift to the overall signal. In IRS-aided antennas, the phase-delayed signal is then reflected from the array whereas in ITS-aided antennas, the phase-delayed signal is transmitted in the forward direction\footnote{We note that IRS/ITS-aided antennas have several advantages/disadvantages with respect to  each other and which one is preferable depends on the particular implementation strategy. For instance, for IRS-aided antennas, the feed position introduces a blocking area whereas this issue does not exist for ITS-aided antennas. On the other hand, IRS facilitates the placement of the control system for the phase shifters on the back side of the surface \cite{abdelrahman2017analysis}. {\BLUE Moreover, for the IRS-aided antennas, the  magnitude of reflection coefficient is often large (close to one) due to the existence of a metal ground plane that reflects the entire incident wave, see Fig.~\ref{Fig:SysMod} e). In contrast, for ITS-aided antennas, the intelligent surface has to be properly designed to ensure a large magnitude for the transmission coefficient which in general may lead to a higher implementation complexity~\cite{abdelrahman2017analysis}.}}. Borrowing an analogy from optics, an IRS is analogous to a curved mirror  whereas an ITS is analogous to a lens. The curvatures of this imagined mirror and lens are steerable via the reconfigurable phase shifters. Unlike in LA antennas, where the EM lens is non-reconfigurable, i.e., non-intelligent, and the direction of the beam is controlled by the location of the corresponding active antenna, in IRS/ITS-aided antennas, the direction of the beam is directly controlled by the reconfigurable intelligent surface. Therefore, the numbers of active antennas in IRS/ITS-aided antennas do not have to scale with the number of passive elements\footnote{LA antennas can be seen as special cases of ITS-aided antennas. In fact, ITS-aided antennas reduce to LA antennas when \textit{i)} the surface is non-reconfigurable and is designed to focus a wavefront arriving perpendicular at the surface to its focal point and \textit{ii)} the active antennas are placed on the focal arc of the surface (i.e., the lens).}. {\BLUE The performance gain of IRS/ITS-aided antennas compared to hybrid antennas is often attributed to the feed mechanism. In particular, the analog network which feeds the transmit antennas in  hybrid architectures causes a severe power loss, which impedes its implementation in massive MIMO and high-frequency systems \cite{abdelrahman2017analysis}. In contrast, the IRS/ITS-aided antennas feed the transmit antennas on the intelligent surface  over the air, which is referred to as space feeding mechanism and inherently more energy efficient \cite{abdelrahman2017analysis,nayeri2018reflectarray}.}

We note that IRS/ITS-aided antennas have been widely investigated in the microwave and antenna community, where they are also known as reflect/transmit arrays \cite{abdelrahman2017analysis,luyen2019wideband}, quasi-optical arrays \cite{popovic1998quasi}, and reconfigurable arrays \cite{di2015reconfigurable,luyen2019wideband}. Moreover, various prototypes are available in the literature \cite{luyen2019wideband,pham2019low,hasani2019dual}. Thereby, the performance of these architectures is typically characterized in terms of the beamforming gain. In contrast, in this paper, we are interested in multiplexing several data streams and the design of the corresponding precoder. In particular, this paper makes the following contributions:
\begin{itemize}
	\item We first model the precoder structure of IRS/ITS-aided antennas and show that the constraints it has to meet are different compared to those for conventional MIMO architectures. In addition, we introduce several illumination strategies (i.e., choices for the positions and orientations of  the active antennas with respect to the intelligent surface) which affect the precoder structure.
	\item Taking the constraints on the precoder into account and exploiting the sparsity of mmWave channels, we design two efficient precoders for IRS/ITS-aided antennas; one based on maximizing the mutual information (MI) and one based on approximating the optimal unconstrained FD precoder via orthogonal matching pursuit (OMP). The performance of the MI-based precoder serves as an upper bound for that of the OMP-based precoder; however, the computational complexity of the former is higher than that of the latter. Hence, using the MI-based precoder as a performance upper bound allows us to assess the efficiency of the OMP-based precoder. 
	\item We develop a power consumption model for IRS/ITS-aided antennas that takes into account the impact of several IRS/ITS imperfections, namely the spillover loss, taper loss, aperture loss, and phase shifter loss, as well as the power consumption of the required digital signal processing and the power amplifiers. For a fair performance comparison, for conventional MIMO architectures, we adopt power consumption models from the literature \cite{yan2018performance,wang2017spectrum} that account for their unique characteristics, e.g., the losses in the RF feed networks of hybrid architectures and the losses in the switching network of the LA architecture. We show that the power consumption of the conventional FD, FC, and PC architectures significantly increases as a function of the number of transmit antennas whereas the power consumption of the LA and IRS/ITS-aided antennas is almost independent of the number of transmit antennas\footnote{Throughout the paper, we refer to the passive elements of the LA and IRS/ITS-aided architectures as transmit antennas, too.}.
	\item We study the impact of the system parameters on the achievable rate via simulations and show that a proper positioning of the active antennas with respect to the intelligent surface in IRS/ITS-aided antennas leads to a considerable performance improvement. In addition, our simulation results show that in contrast to the conventional FD, FC, PC, and LA MIMO architectures,  the IRS/ITS-aided MIMO architectures are fully scalable in terms of the number of transmit antennas\footnote{We note that, at mmWave frequencies, hundreds and even thousands of passive antenna elements can be accommodated in a compact design. For example, a 10~cm-by-10~cm intelligent surface can contain approximately 350 and 1600 passive elements at frequencies of $28$~GHz and $60$~GHz, respectively, with element spacing of half a wavelength.}. Therefore, IRS/ITS-aided antennas are promising candidates for realizing the potential of mmWave \textit{ultra  massive} MIMO~in~practice.
\end{itemize}

We note that the recent paper \cite{zhou2018hardware} also studied IRS-aided antennas (referred to as reflect arrays) and proposed a corresponding precoder design based on alternating optimization (AO). We employ this precoder as a benchmark and show that the proposed precoders outperform the AO-based precoder in \cite{zhou2018hardware} especially for environments with few scatterers. Moreover, the focus of this paper is mainly on the scalability and energy efficiency of IRS/ITS-aided antennas which were not studied in \cite{zhou2018hardware}. Furthermore, compared to \cite{zhou2018hardware}, in this paper, more detailed models for the precoder structure and the power consumption of the IRS/ITS-aided antennas are provided. Such accurate models are needed for a fair performance comparison of different MIMO architectures and the design of IRS/ITS-aided MIMO structures (e.g. the adjustment of the relative positions and orientations of the active antennas with respect to the intelligent surface) which was not investigated~in~\cite{zhou2018hardware}.  

In our previous work \cite{bereyhi2019papr,bereyhi2020single}, we investigated symbol-level precoding of intelligent surfaces where the update rate was equal to the symbol rate. In contrast, in this paper, we design the precoder for a given channel realization, which implies that the state of the intelligent surface has to be updated once per channel coherence time. Therefore, the complexity of the IRS/ITS-aided antennas considered in this paper is lower than that of the symbol-level precoding schemes studied in  \cite{bereyhi2019papr,bereyhi2020single}.   Finally, we note that this paper extends its conference version  \cite{jamali2018scalable} in the following directions: \textit{i)} The precoder structure for the IRS/ITS-aided antenna architectures is formulated more carefully in this paper (see assumptions A1-A4 in Section~II-A) and the special case of  hypothetical uniform illumination is considered, cf. Corollary~1, which is not included in \cite{jamali2018scalable}. \textit{ii)} The power consumption of the IRS/ITS-aided antenna architectures is discussed in more detail via an example setup in Section~III-D. {\BLUE Moreover, we establish  lower and upper bounds on the power radiated by the active antennas in terms of the desired power to be radiated by the intelligent surface into the channel, cf. Lemma~\ref{Lem:PrdPtx}, which is not considered in \cite{jamali2018scalable}.}  \textit{iii)} In this paper, we propose two precoder designs, namely MI-based and OMP-based precoders, whereas in \cite{jamali2018scalable}, only the OMP-based precoder is given. \textit{iv)}  We include the LA architecture as a new benchmark scheme in this paper and provide  corresponding models for the precoder structure and power consumption using the same unified framework as for the other benchmark architectures, see Table~II. 

The rest of this paper is organized as follows. In Section~\ref{Sec:SigSysModel}, we provide the considered system, channel, and signal models. Mathematical models for the precoder structure and the power consumption of the considered IRS/ITS-aided antenna architectures are presented in Section~\ref{Sec:RA_TA}.  In Section~\ref{Sec:Precoder}, two different precoder designs for IRS/ITS-aided antennas are developed. Simulation results are provided in Section~\ref{Sec:Sim}, and conclusions and directions for future research are presented in Section~\ref{Sec:Cncln}.

\textit{Notations:} Bold capital and small letters are used to denote matrices and vectors, respectively. $\|\mathbf{A}\|_F$, $\Tr(\mathbf{A})$, $\mathbf{A}\Trans$, and $\mathbf{A}\Herm$ denote the Frobenius norm, trace, transpose, and Hermitian  of matrix $\mathbf{A}$, respectively. $\Ex\{\cdot\}$ represents expectation and $I(\mathbf{x};\mathbf{y})$ denotes the MI between random variables (RVs) $\mathbf{x}$ and $\mathbf{y}$. $|a|$ and $\angle a$ denote the absolute value and the angle of complex number $a$ in polar coordinates, respectively. In addition, $|\mathbf{A}|$ denotes the determinant of square matrix $\mathbf{A}$.  The big O notation, $g(x)=O(f(x))$, indicates $\lim_{x\to\infty}|g(x)/f(x)|\leq k$ for some fixed $k$, where $0<k<\infty$. For a real number $x$, $[x]^+=\max\{0,x\}$ and $y=\lceil x \rceil$ is the smallest integer number $y$ for which $y\geq x$ holds.  $\mathcal{CN}(\boldsymbol{\mu},\boldsymbol{\Sigma})$ denotes a complex Gaussian RV with mean vector $\boldsymbol{\mu}$ and covariance matrix $\boldsymbol{\Sigma}$. Furthermore, $\boldsymbol{0}_n$ and $\boldsymbol{0}_{n\times m}$ denote a vector of size $n$ and a matrix of size $n\times m$, respectively, whose elements are all zeros. $\mathbf{I}_n$ is the $n\times n$ identity matrix and $\mathbb{C}$ represents the set of complex numbers. $[a(m,n)]_{m,n}$ represents a matrix with element $a(m,n)$ in its $m$-th row and $n$-th column. $\mathbf{A}_{m,n}$ and $\mathbf{a}_n$ denote the element in the $m$-th row and $n$-th column of matrix $\mathbf{A}$ and the $n$-th element of vector $\mathbf{a}$, respectively. Finally, $\mathrm{vec}(\mathbf{A})$ denotes a vector whose elements are the stacked columns of matrix~$\mathbf{A}$, {\BLUE and $\otimes$ denotes the Kronecker product}.

\section{Signal, System, and Channel Models}\label{Sec:SigSysModel}
In this section, we present the system, transmit signal, and channel models for the considered IRS/ITS-aided MIMO systems.

\subsection{System Architecture}\label{Sec:Sys}

\begin{figure}
	\centering
	\includegraphics[width=1\linewidth]{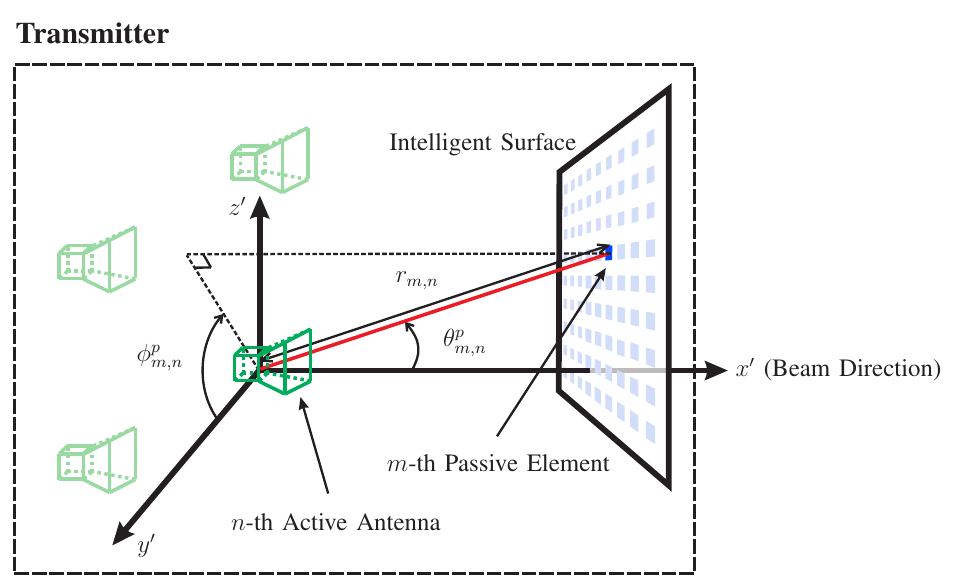} \vspace{-0.3cm} 
	\caption{\BLUE Schematic illustration of positioning of the active antennas and the intelligent surface. \vspace{-0.03cm}}
	\label{Fig:Positioning}
\end{figure}

We assume that the considered IRS/ITS-aided antennas are equipped with $N$ active antennas and that the intelligent surface comprises $M$ passive antenna elements. Moreover, we assume that each active feed antenna is connected to a dedicated RF chain, i.e., there are $N$ RF chains. {\BLUE  To facilitate presentation, we characterize the positions of the passive antenna elements by $(r_{m,n},\theta_{m,n}^p,\phi_{m,n}^p),\,\,r_{m,n}\geq 0,\theta^p_{m,n}\in[0,\frac{\pi}{2}],\phi^p_{m,n}\in[0,2\pi]$, in $N$ different spherical coordinate systems whose respective origins are the positions of the active antennas. Here,   $\theta^p_{m,n}$ denotes the elevation angle of passive element $m$ (with respect to the beam direction of active antenna $n$), $\phi^p_{m,n}$ represents the azimuth angle of passive element $m$ (in the plane perpendicular to the beam direction of active antenna $n$), and $r_{m,n}$ is the distance between  passive element $m$ and active antenna $n$, see Fig.~\ref{Fig:Positioning} for an illustration of $(r_{m,n},\theta^p_{m,n},\phi^p_{m,n})$. In a similar manner, we characterize the positions of the active antenna elements by $(r_{m,n},\theta_{m,n}^a,\phi_{m,n}^a),\,\,r_{m,n}\geq 0,\theta^a_{m,n}\in[0,\frac{\pi}{2}],\phi^a_{m,n}\in[0,2\pi]$, in $M$ different spherical coordinate systems whose respective origins are the positions of the passive antennas. Here, $\theta^a_{m,n}$ denotes the elevation angle of active element $n$ (with respect to the normal to the surface) and $\phi^a_{m,n}$ represents the azimuth angle of active element $n$ (in the surface plane).}  Note that the values of $(r_{m,n},\theta^p_{m,n},\phi^p_{m,n})$ {\BLUE and $(r_{m,n},\theta^a_{m,n},\phi^a_{m,n})$}  depend on the specific positioning of the feed antennas and the intelligent surface. Moreover, in order to rigorously present our results, we make the following assumptions. 
\begin{itemize}
	\item[A1)] We assume the same antenna pattern for all active antennas with antenna gain $G^a(\theta,\phi)$ for elevation angle $\theta$ and azimuth angle $\phi$ {\BLUE defined in the spherical coordinate system with the active antenna as the origin. Similarly, we assume  the same antenna pattern for all passive antennas with antenna gain $G^p(\theta,\phi)$ for elevation angle $\theta$ and azimuth angle $\phi$ defined  in the spherical coordinate system with the passive antenna as the origin.} 
	\item[A2)]  We assume that while each element of the intelligent surface is in the far field of the active antennas; the entire surface is not in the far field. This implies that the electric-field power reaching each element decays with the distance square $\propto\frac{1}{r_{m,n}^2}$ (i.e., higher order terms $\propto\frac{1}{r_{m,n}^q},\,\,q>2$, are negligible); however, the wavefront phase curvature cannot be neglected. These are  valid assumptions if $\lambda \ll r_{m,n}\leq 2d^2_{e}\lambda,\,\,\forall m,n$ (typically $r_{m,n}\geq 5 \lambda,\forall m,n$), which holds in practice \cite{balanis1982antenna}. Here, $\lambda$ and $d_{e}$ denote the wavelength and the largest electric dimension (defined as the physical dimension normalized to $\lambda$) of the intelligent surface, respectively.
	\item[A3)] We neglect the mutual coupling between the active antennas (passive antenna elements) which is an accurate assumption when the antennas (passive elements) are sufficiently separated, i.e., typically by at least $\lambda/2$  \cite{masouros2013large}. 
	\item[A4)] We assume that the power radiated from the active antennas is either reflected/forwarded or absorbed by the intelligent surface such that no power from the active antennas directly arrives at the receiver. 
\end{itemize}

%


\subsection{Transmit Signal Model}\label{Sec:Signal}

Let $\mathbf{s}\in\mathbb{C}^{Q\times 1}$ denote the vector of $Q$ independent data streams that we wish to transmit. Moreover, let $\mathbf{x}\in\mathbb{C}^{M\times 1}$ denote the transmit  vector radiated from the intelligent surface. Assuming linear precoding, the relation between transmit vector $\mathbf{x}$ and  data vector $\mathbf{s}$ is as follows
\begin{IEEEeqnarray}{lll} \label{Eq:Sig_Gen}
	\mathbf{x} =\sqrt{\Ptx}\mathbf{F}\mathbf{s},
\end{IEEEeqnarray}
where $\mathbf{F}\in\mathbb{C}^{M\times Q}$ is the precoder which includes the baseband precoder, the  impact of the channel between the active antennas and the intelligent surface, the imperfections of the IRS/ITS, and the phase change introduced by the intelligent surface, see Section~\ref{Sec:RA_TA} for the detailed modeling of precoder $\mathbf{F}$. Moreover, in \eqref{Eq:Sig_Gen}, $\Ptx$ denotes the transmit power radiated by the intelligent surface and we assume that $\Ex\{\mathbf{s}\mathbf{s}\Herm\}=\mathbf{I}_Q$ and $\|\mathbf{F}\|_F = 1$ hold.  In this paper, we impose a constraint on the maximum power radiated from the intelligent surface into the channel which is typically enforced by regulations. For example, for carrier frequencies $54$-$66$ GHz, the United States Federal Communications Commission enforces a total maximum transmit power of $500$ mW ($27$ dBm) for an emission bandwidth of more than $100$ MHz \cite{federal2016use}. Alternatively, one can impose a constraint on the power radiated by the active antennas\footnote{The maximum radiated power is also constrained by the antenna effective isotropic radiated power (EIRP) \cite{federal2016use}. The transmit power and EIRP are related according to $\text{EIRP}=G_{\max}\Ptx$, where $G_{\max}$ is the overall maximum antenna gain of the passive elements on the intelligent surface.}. Although our derivations in Section~\ref{Sec:RA_TA} and the proposed precoders in Section~\ref{Sec:Precoder} are applicable for both power constraints, we focus on the former power constraint for IRS/ITS-aided antennas since this enables a more straightforward comparison with conventional MIMO architectures.

\subsection{Channel Model}\label{Sec:SysMod}
\begin{figure}
	\centering 
	\includegraphics[width=0.7\linewidth]{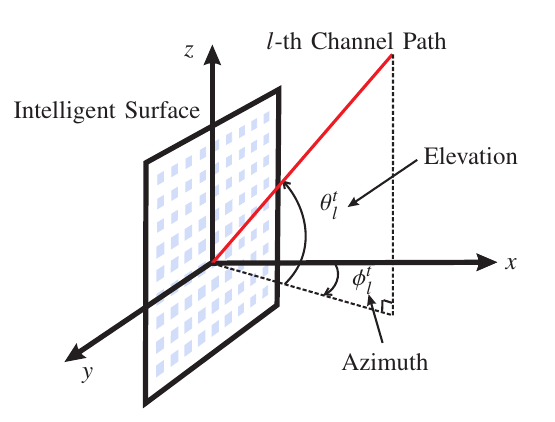} \vspace{-0.1cm} 
	\caption{Schematic illustration of a given channel AoD with respect to the intelligent surface. \vspace{-0.03cm}}
	\label{Fig:SysModAnglesys}
\end{figure}

We consider a point-to-point MIMO system with the following input-output channel model 
\begin{IEEEeqnarray}{lll} \label{Eq:AWGN}
\mathbf{y} = \Hc\mathbf{x}+\mathbf{z},
\end{IEEEeqnarray}
where $\mathbf{y}\in\mathbb{C}^{J\times 1}$ denotes receive vector and $J$ is the number of receive antennas. Moreover, $\mathbf{z}\in\mathbb{C}^{J\times 1}$ denotes the additive white Gaussian noise vector at the receiver, i.e., $\mathbf{z}\sim\mathcal{CN}(\boldsymbol{0}_J,\sigma^2\mathbf{I}_J)$ where $\sigma^2$ is the noise variance at each receive antenna. Furthermore, $\Hc\in\mathbb{C}^{J\times M}$  is a low-rank channel matrix accounting for the limited number of scatterers in the channel. In particular, $\Hc$ is modeled as follows \cite{busari2017millimeter,delgado2018feasible,akdeniz2014millimeter}
\begin{IEEEeqnarray}{lll} \label{Eq:Channel}
\Hc = \frac{1}{\sqrt{L}}\sum_{l=1}^L h_l \mathbf{h}_r(\theta_l^r,\phi_l^r)\mathbf{h}_t\Herm(\theta_l^t,\phi_l^t).
\end{IEEEeqnarray}
where $L$ is the number of effective channel paths  and $h_l\in\mathbb{C}$ is the channel coefficient of the $l$-th path. Moreover, $\mathbf{h}_t(\theta_l^t,\phi_l^t)$ ($\mathbf{h}_r(\theta_l^r,\phi_l^r)$) denotes the transmitter (receiver) antenna array steering vector for angle-of-departure (AoD) $(\theta_l^t,\phi_l^t)$ (angle-of-arrival (AoA) $(\theta_l^r,\phi_l^r)$) with elevation angle $\theta_l^t\in[-\pi/2,\pi/2]$ ($\theta_l^r\in[-\pi/2,\pi/2]$) and azimuth angle $\phi_l^t\in[-\pi/2,\pi/2]$ ($\phi_l^r\in[-\pi/2,\pi/2]$)\footnote{We note that for notational convenience, in this paper, the definitions of elevation and azimuth angles used for characterization of $(r_{m,n},\theta^p_{m,n},\phi^p_{m,n})$ and $(r_{m,n},\theta^a_{m,n},\phi^a_{m,n})$ are different from those used for characterization of the AoAs/AoDs of the channel, see Figs. \ref{Fig:Positioning} and \ref{Fig:SysModAnglesys}. The latter follows the standard definition of spherical coordinate systems in the physics literature \cite[Chapter 16]{balanis1982antenna} whereas the former is a popular convention used in the radar literature  \cite{zeng2018multi}.}, see Fig.~\ref{Fig:SysModAnglesys}. Assuming the passive elements are uniformly distributed on the intelligent surface which itself lies in the $y-z$ plane,  $\mathbf{h}_t(\phi_l^t,\theta_l^t)$ is given as follows  \cite{busari2017millimeter}
\begin{IEEEeqnarray}{lll} \label{Eq:UPA}
	\mathbf{h}_t(\theta_l^t,\phi_l^t)  \nonumber\\
	\!=\! 
	\mathrm{vec}\left(\left[e^{j\frac{2\pi d}{\lambda}\left((m_y-1)\cos(\theta_l^t)\sin(\phi_l^t)+(m_z-1)\sin(\theta_l^t)\right)}\right]_{m_y,m_z}\right) \nonumber\\
	\BLUE=\mathrm{vec}\left(\mathbf{h}_t(\theta_l^t) \otimes \mathbf{h}_t^{\mathsf{T}}(\phi_l^t|\theta_l^t) \right),\quad
\end{IEEEeqnarray}
{\BLUE where 
\begin{IEEEeqnarray}{lll} \label{Eq:UPA_vector}
	\mathbf{h}_t(\theta_l^t) = \left[e^{j\frac{2\pi d}{\lambda}(m_z-1)\sin(\theta_l^t)}\right]_{m_z} 
	\IEEEyesnumber\IEEEyessubnumber\\
	\mathbf{h}_t(\phi_l^t|\theta_l^t) = \left[e^{j\frac{2\pi d}{\lambda}(m_y-1)\cos(\theta_l^t)\sin(\phi_l^t)}\right]_{m_y}.\IEEEyessubnumber
\end{IEEEeqnarray}
Here, $d$ is the distance between the passive antenna elements. Moreover, assuming a square surface and that $\sqrt{M}$ is integer, the passive antenna indices along the $y$- and $z$-axes are denoted by $m_y,m_z=1,\dots,\sqrt{M}$. Furthermore, as shown in \eqref{Eq:UPA}, the considered steering vector can be decomposed into the Kronecker product of two vectors, namely $\mathbf{h}_t(\theta_l^t)$ and $\mathbf{h}_t(\phi_l^t|\theta_l^t)$ which show the  capability of the array to scan different elevation angles and different azimuth angles for a given elevation angle, respectively.} We assume an FD receiver equipped with a uniform planar antenna array. Thus, $\mathbf{h}_r(\theta_l^r,\phi_l^r)$ is defined analogous to $\mathbf{h}_t(\theta_l^t,\phi_l^t)$ in (\ref{Eq:UPA}).

%
%

\section{Mathematical Modeling of the IRS/ITS-Aided MIMO Architectures}\label{Sec:RA_TA}

In this section, we first model the constraints that the IRS/ITS-aided antennas impose on the precoder.  Then, we present different illumination strategies for the active antennas. Finally, we quantify the total power consumption of IRS/ITS-aided MIMO and further elaborate on the inherent losses for a  special~case.

\subsection{Constraints on the Precoder} 

 Let us define $\bar{\mathbf{x}}=[\bar{x}_1,\dots,\bar{x}_N]\Trans\in\mathbb{C}^{N\times 1}$ and $\bar{\mathbf{y}}=[\bar{y}_1,\dots,\bar{y}_M]\Trans\in\mathbb{C}^{M\times 1}$ where $\bar{x}_n$ and $\bar{y}_m$ denote the signal transmitted by the $n$-th active antenna and the signal received at the $m$-th passive antenna, respectively. As can be seen from Fig.~\ref{Fig:SysMod}, the data stream vector $\mathbf{s}$ is multiplied by the baseband precoder $\mathbf{B}$, fed to the RF chains, and then transmitted over the active antennas/illuminators, i.e., $\bar{\mathbf{x}}=\sqrt{\Ptx}\mathbf{B}\mathbf{s}$. {\BLUE We note that the maximum number of independent data streams $Q$ that can be supported is limited by both the numbers of transmitter RF chains $N$ and the rank of the channel matrix $\min\{M,J,L\}$. Moreover,  the numbers of transmit and receive antennas are typically large in mmWave communication systems in order to ensure a sufficient link budget, i.e., $M,J\gg N,L$, which implies that for the maximum number of independent data streams, we typically have $Q\leq \min\{N,L\}$.} Based on assumptions A1-A3, the signal that is received at the $m$-th passive element, $\bar{y}_{m}$, is obtained as \cite{di2015reconfigurable}
\begin{IEEEeqnarray}{lll} \label{Eq:Sig}
	\bar{y}_{m} \!=\! \sum_{n=1}^N \!\sqrt{\!{\BLUE G^a(\theta^p_{m,n},\phi^p_{m,n})G^p(\theta^a_{m,n},\phi^a_{m,n})}} \frac{\lambda e^{-j\frac{2\pi r_{m,n}}{\lambda}}}{4\pi r_{m,n}} \bar{x}_n.\quad
\end{IEEEeqnarray}
The intelligent surface applies a phase shift of $2\pi\beta_m$ to the  signal received at the $m$-th element before reflecting/transmitting it, i.e., $x_m=\bar{y}_m \sqrt{\rho_{\mathrm{srf}}}\exp(j2\pi \beta_m)$, where $x_m$ is the $m$-th element of $\mathbf{x}$. The signal attenuation caused by the aperture  and phase shifter losses is captured by a power efficiency factor denoted by $\rho_{\mathrm{srf}}$, see Section~\ref{Sec:Loss_RA_TA} for details. The following proposition formally characterizes the precoder structure for IRS/ITS-aided antennas in matrix~form.

\begin{prop}\label{Prop:Precoder}
	Under Assumptions A1-A4, the precoder $\mathbf{F}$ for IRS/ITS-aided antennas has the form
	\begin{IEEEeqnarray}{lll}\label{Eq:Precoder} 
		\mathbf{F} = \mathbf{D}\mathbf{T}\mathbf{B},
	\end{IEEEeqnarray}
	where  $\mathbf{B}\in\mathbb{C}^{N\times Q}$ is the digital baseband precoder which controls the output signal of the active antennas. $\mathbf{D}\in\mathbb{C}^{M\times M}$  is a diagonal phase-shift matrix which controls the intelligent surface and is given by
	\begin{IEEEeqnarray}{lll}\label{Eq:Dmatrix} 
		\mathbf{D}=\Diag\left(e^{j2\pi\beta_1},\dots,e^{j2\pi\beta_M}\right)
	\end{IEEEeqnarray}
	with $\beta_m\in[0,1]$. Moreover, $\mathbf{T}\in\mathbb{C}^{M\times N}$ is a fixed matrix, which depends on the power efficiency of the intelligent surface, denoted by $\rho_{\mathrm{srf}}$, active and passive antenna gains, and the antenna positioning, i.e., {\BLUE $(r_{m,n},\theta^p_{m,n},\phi^p_{m,n})$ and $(r_{m,n},\theta^a_{m,n},\phi^a_{m,n}),\,\,\forall m,n$}, and is given~by
	\begin{IEEEeqnarray}{lll} \label{Eq:Tmaxtrix}
		\mathbf{T}\!=\!\left[\frac{\lambda\sqrt{\rho_{\mathrm{srf}}{\BLUE G^a(\theta^p_{m,n},\phi^p_{m,n})G^p(\theta^a_{m,n},\phi^a_{m,n})}}}{4\pi r_{m,n}} e^{-j\frac{2\pi r_{m,n}}{\lambda}} \right]_{m,n} 
		\hspace{-0.6cm}.\quad\,\,
	\end{IEEEeqnarray}
\end{prop}
\begin{IEEEproof}
 The proof  follows from rewriting \eqref{Eq:Sig} in matrix form $\bar{\mathbf{y}}=\mathbf{T}\bar{\mathbf{x}}$, where  the  efficiency factor $\rho_{\mathrm{srf}}$ is included in $\mathbf{T}$, and combining it with  $\mathbf{x}=\mathbf{D}\bar{\mathbf{y}}$, $\bar{\mathbf{x}}=\sqrt{\Ptx}\mathbf{B}\mathbf{s}$, and the definition of the linear precoder $	\mathbf{x} =\sqrt{\Ptx}\mathbf{F}\mathbf{s}$ in \eqref{Eq:Sig_Gen}.
\end{IEEEproof}

We note that both IRS/ITS-aided antennas have identical precoder structures, as given by (\ref{Eq:Precoder}). The precoder for IRS/ITS-aided antennas consists of three parts, namely $\mathbf{B}$, $\mathbf{T}$, and $\mathbf{D}$. Among these three components, $\mathbf{T}$ is fixed and determined during manufacturing whereas $\mathbf{B}$ and $\mathbf{D}$ can be adjusted during online transmission based on the CSI. In Section~\ref{Sec:Illumination}, we introduce different strategies for the design of matrix $\mathbf{T}$. Moreover, in Section~\ref{Sec:Precoder}, we propose two precoding schemes for optimization of matrices $\mathbf{B}$ and $\mathbf{D}$.  We note that $\mathbf{T}$ and particularly the surface power efficiency factor $\rho_{\mathrm{srf}}$ may assume different values for IRS/ITS-aided antennas, respectively, see Section~\ref{Sec:Loss_RA_TA} for details.

\subsection{Illumination Strategies}\label{Sec:Illumination}

\begin{figure*}
	\centering
	\includegraphics[width=0.8\linewidth]{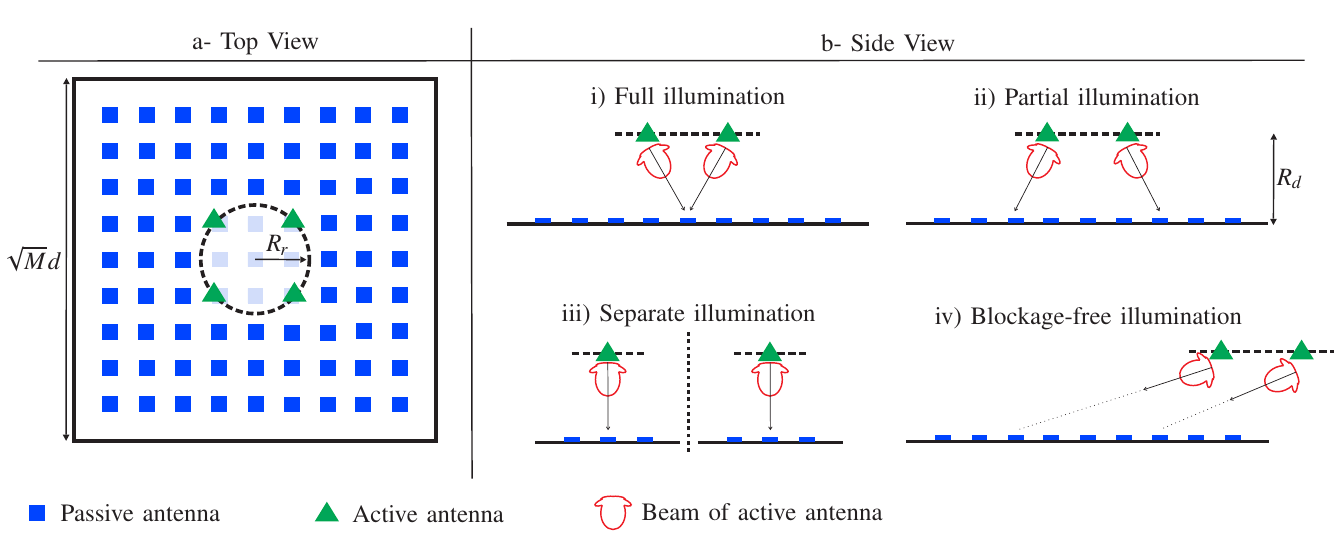}\vspace{-0.02cm} 
	\caption{Schematic illustration of the arrangement of the active antennas and the intelligent surface for different illumination strategies.  \vspace{-0.93cm}}
	\label{Fig:Illumination}
\end{figure*}

Matrix $\mathbf{T}$ depends on how the active antennas illuminate the intelligent surface of the IRS/ITS-aided antennas. Thus, different illumination strategies, including the relative positioning and orientation of the active antennas and the intelligent surface, lead to different designs of matrix $\mathbf{T}$. In the following, we introduce several different illumination strategies. To do so, let us assume that the intelligent surface lies in the $y-z$ plane and its center is at the origin. Furthermore, let us assume that all active antennas have distance $R_d$ from the intelligent surface and are uniformly distributed on a ring of radius $R_r$ with center $\mathbf{c}$, see Fig.~\ref{Fig:Illumination}. We assume this specific geometry for the locations of the active antennas and the intelligent surface in order to be able to rigorously present the proposed illumination strategies. Nevertheless, we note that other geometries are also possible and only affect matrix $\mathbf{T}$ while the precoder designs proposed in Section~\ref{Sec:Precoder} are valid for any given matrix $\mathbf{T}$.

\textbf{Full Illumination (FI):} Here, each active antenna fully illuminates the entire intelligent surface \cite{arrebola2008multifed,zhou2018hardware}. To achieve this, we assume that all active antennas illuminate the center of the intelligent surface, see Fig.~\ref{Fig:Illumination}~b-i).

\textbf{Partial Illumination (PI):} Note that each passive element can only change the phase of the \textit{superposition} of the signals that it receives from the active antennas. To compensate for this limitation, a natural option is to have the active antennas illuminate disjoint subsets of the intelligent surface. Assuming that the passive  elements responsible for a given active antenna are	physical neighbors, for PI, each active antenna illuminates the	center of the area occupied by its respective passive elements, see Fig.~\ref{Fig:Illumination}~b-ii).

\textbf{Separate Illumination (SI):} SI is an extreme special case of PI where the signals of different active antennas are physically shielded such that each part of the intelligent surface only receives the signal of one active antenna, see Fig.~\ref{Fig:Illumination}~b-iii). Note that due to the wide beam patterns of the active antennas, even under PI, they illuminate not only their respective subset of passive elements, but all other passive elements as well. 	This causes interference between the signals from different RF chains which is avoided~by~SI.

\textbf{Blockage-Free PI:} In this paper, in general, for the IRS-aided architecture, we neglect that the active antennas may partially block the reflected RF signal. Therefore, unless otherwise stated, we assume $\mathbf{c}=(R_d,0,0)$ which yields the minimum average distance between the active  antennas and passive elements, i.e., $\frac{1}{MN}\sum_{n}\sum_{m}r_{m,n}$ and thus the over-the-air losses are minimized. In practice, IRS-aided antennas are designed to support transmission/reception for a limited range of AoDs/AoAs and the active antennas are placed outside this range to avoid blockage \cite{arrebola2008multifed,pozar1997design}. To achieve this, for blockage-free illumination, we assume that the active antennas are placed on one side of the intelligent surface (e.g., $\mathbf{c}=(R_d,d\sqrt{M}/2,d\sqrt{M}/2)$) in order to avoid blocking the AoAs/AoDs pointing to the other side \cite{arrebola2008multifed,pozar1997design}, see Fig.~\ref{Fig:Illumination}~b-iv). For simplicity, in this paper, we use blockage-free illumination only in combination with PI although in principle it can be also used in combination with FI and SI, e.g., see  \cite{arrebola2008multifed} for blockage-free FI.

\textbf{Hypothetical Uniform SI:} For practical illumination strategies, the distribution of the power received from each active antenna at the intelligent surface is non-uniform. For single-stream data transmission, non-uniform power distribution causes a decrease in the achievable antenna gain, which is known as taper loss \cite[Chapter~15]{balanis1982antenna}. More generally, for multi-stream data transmission, a non-uniform power distribution reduces the achievable rate. In order to study the performance degradation due to non-uniform power distribution across the intelligent surface, we focus on SI and  introduce a hypothetical illumination where the powers of the signals received from each active antenna at its respective passive elements are identical. To formally model uniform SI, let us rewrite matrix $\mathbf{T}=\big[c_{m,n}e^{-j2\pi r_{m,n}/\lambda}\big]_{m,n}$, where $c_{m,n}=\lambda\sqrt{\rho_{\mathrm{srf}}{\BLUE G^a(\theta^p_{m,n},\phi^p_{m,n})G^p(\theta^a_{m,n},\phi^a_{m,n})}}/(4\pi r_{m,n})$. For uniform SI, the phase $2\pi r_{m,n}/\lambda$ remains the same as for SI while $c_{m,n}$ is equal to a constant $c$, $\forall n,m\in\mathcal{M}_n$, and equal to zero otherwise. Here,  $\mathcal{M}_n$ is the set of indices of the passive elements responsible for reflection/transmission of the signal emitted by active antenna $n$. To account for the over-the-air pathloss and surface efficiency, we set the value of $c$ as $\lambda\sqrt{\rho_{\mathrm{srf}}{\BLUE G^aG^p}}/(4\pi r)$, where $r=\frac{1}{M}\sum_{n}\sum_{m\in\mathcal{M}_n}r_{m,n}$ and {\BLUE we assume $G^a(\theta^p_{m,n},\phi^p_{m,n})\triangleq G^a,\,\,\forall \theta^p_{m,n}\in[0,\theta_0^{\mathrm{SI}}],\forall\phi^p_{m,n}\in[0,2\pi]$, and $G^p(\theta^a_{m,n},\phi^a_{m,n})\triangleq G^p,\,\,\forall \theta^a_{m,n}\in[0,\frac{\pi}{2}],\forall\phi^p_{m,n}\in[0,2\pi]$\footnote{\BLUE For  the hypothetical uniform SI, for the active antennas, we assume a uniform antenna pattern with the elevation angle confined to $\theta\in[0,\theta_0^{\mathrm{SI}}]$, whereas for the passive antennas, we assume a uniform antenna pattern with the elevation angle confined to $\theta\in[0,\frac{\pi}{2}]$. The reason for this choice is that the active antennas are usually horn antennas whose antenna gain is controllable, whereas the passive antennas are typically simple patch antennas whose antenna pattern is wide and not easy to control.}.} Here, $G^a$ and $G^p$ are constants and $\theta_0^{\mathrm{SI}}$ is the elevation angular extent of the sub-surfaces in SI with respect to their feed antennas.  The following corollary provides matrix $\mathbf{T}$ for uniform SI. 

\begin{corol}\label{Corol:Ideal}
	For uniform SI, matrix $\mathbf{T}$ in \eqref{Eq:Tmaxtrix} simplifies to
	\begin{IEEEeqnarray}{lll}\label{Eq:PrecoderSimp} 
		\mathbf{T}=\left[c \, e^{-j\frac{2\pi r_{m,n}}{\lambda}}\right]_{m,n} 
	\end{IEEEeqnarray}
	with
	\begin{IEEEeqnarray}{lll} 
		c = \begin{cases}
			\frac{\lambda}{4\pi r}\sqrt{\frac{4\rho_{\mathrm{srf}}}{1-\cos(\theta_0^{\mathrm{SI}})}},\hspace{-2mm}&\mathrm{if} \,\,m\in\mathcal{M}_n\\
			0, &\mathrm{otherwise.}
		\end{cases}\quad
	\end{IEEEeqnarray}
	
\end{corol}
\begin{IEEEproof}
	The proof follows directly from noting that {\BLUE $G^a=\frac{2}{1-\cos(\theta_0^{\mathrm{SI}})}$ and $G^p=2$  has to hold such that $\int_{\Omega} \frac{1}{4\pi} G^{q}(\phi,\theta)\mathrm{d}\Omega=1,\,\,q\in\{a,p\}$}, holds where $\mathrm{d}\Omega=\sin(\theta)\mathrm{d}\theta\mathrm{d}\phi$  \cite{balanis1982antenna} and applying the simplifying assumptions of uniform SI in (\ref{Eq:Tmaxtrix}). 
\end{IEEEproof}

In Section~\ref{Sec:Sim}, we comprehensively study the aforementioned illumination strategies via simulations in order to obtain insights regarding their impact on performance for system design.

\subsection{Power Consumption and Losses}\label{Sec:Loss_RA_TA}

The power consumption of IRS/ITS-aided antennas can be divided into two parts:

\textbf{Baseline Circuitry:} The circuit power consumption comprises the power consumed for digital baseband processing, denoted by $\Pbb$, and by each RF chain (including the digital-to-analog converter, local oscillator, and mixer), denoted by $\Prfc$. Although, in principle, $\Pbb$ may depend on $M$ and $N$, in the remainder of this paper, we assume  $\Pbb$ is constant since its impact is typically much smaller than that of $\Prfc$ \cite{fischer2007next,lin2016energy}.

\textbf{Power Amplifier:} The power consumed by the power amplifier (PA) is commonly modeled as $\Prd/\rho_{\mathrm{pa}}$ where $\Prd$ is the  output power radiated by the active antennas and  $\rho_{\mathrm{pa}}$ denotes the power amplifier efficiency \cite{yan2018performance,ribeiro2018energy,Rodriguez2016_MIMO_Loss,lin2016energy}. The  power radiated by the active antennas is given by $\Prd=\Ex\{\bar{\mathbf{x}}\Herm\bar{\mathbf{x}}\}=\Ptx\|\mathbf{B}\|_F^2$. Due to the losses incurred in the channel between the active antennas and the intelligent surface as well as the inefficiencies of the intelligent surface, $\Ptx\leq\Prd$ holds. The main sources of the power loss for IRS/ITS-aided antennas are provided in the following:  

\begin{itemize}
	\item \textbf{Spillover loss:} Since the effective area of the intelligent surface is finite,  some of the power radiated by the active antennas will not be captured by the passive antennas, resulting in a spillover loss \cite{pozar1997design}. We define the efficiency factor $\rho_{S}$ to take the spillover into account. {\BLUE We note that the value of $\rho_{S}$ depends on the relative positions of the active antennas and the intelligent surface, i.e., $(r_{m,n},\theta^p_{m,n},\phi^p_{m,n}),\,\,\forall m,n$, and the radiation pattern of the active antennas, i.e., $G^a(\theta^p_{m,n},\phi^p_{m,n})$. Hence, the impact of  $\rho_{S}$ is implicitly captured by matrix $\mathbf{T}$.}
	
	\item \textbf{Taper loss:} In general, the density of the received power differs across the intelligent surface as it depends on $G^a(\theta^p_{m,n},\phi^p_{m,n})$ and $r_{m,n}$. As discussed earlier, for multi-stream transmission, taper loss leads to a  reduction of the achievable rate. We define the efficiency factor $\rho_{T}$ to account for this loss.  {\BLUE Similar to $\rho_S$, the impact of $\rho_T$ is implicitly captured by matrix $\mathbf{T}$.}
	
	\item \textbf{Aperture loss:}  Ideally, for IRS/ITS-aided antennas, the total power captured by the aperture will be reflected/transmitted.  In practice, however, the actual power transmitted into the wireless channel is smaller than the  total  power captured by the surface, which is due to inefficiencies in antenna reception/transmission, power absorption by the IRS/ITS, and unwanted reflection by the ITS. The aperture efficiency is taken into account by introducing  efficiency factor $\rho_{A}$. {\BLUE For future reference, we decompose the aperture efficiency factor as $\rho_{A}=\rho_{A}^{\rm ant}\rho_{A}^{\rm srf}$, where $\rho_{A}^{\rm ant}$ is the antenna aperture efficiency of the individual passive antennas and $\rho_{A}^{\rm srf}$ denotes the surface efficiency factor which captures the remaining losses of the surface excluding the antenna aperture losses. Thereby, the impact of $\rho_{A}^{\rm ant}$ is implicitly captured via the passive antenna pattern $G^p(\theta^a_{m,n},\phi^a_{m,n})$ by  matrix~$\mathbf{T}$.}
	
	\item \textbf{Phase shifters:} Each phase shifter introduces a certain loss once the RF signal passes through it.  For the IRS-aided antennas, the signal passes twice through the phase-shifters whereas for the ITS-aided antennas, the signal passes only once through the phase-shifter, {\BLUE  see the dashed arrows in Fig.~\ref{Fig:SysMod}e) and f). Hence, for the reflected wave in the IRS-aided antennas and the re-transmitted wave in the ITS-aided antennas to have the same overall phase shift, the phase shift induced by the phase shifters of the IRS-aided antennas should be half of that for the ITS-aided antennas. The impact of this difference between IRS- and ITS-aided antennas on the overall phase-shifter losses depends on the specific phase-shifter technology. Let $\rho_P(\beta)$ denote the phase-shifter power efficiency factor as a function of phase-shift value $\beta$. Then, the overall phase shifter efficiency factors of the IRS- and ITS-aided antennas are given by $\rho_P^2(\beta/2)$ and $\rho_P(\beta)$, respectively. In practice, for most phase-shifter realizations, the overall phase-shifter loss is almost independent of the phase-shift value and practically constant \cite{hum2013reconfigurable,maloratsky2010electrically}. For example,  in switched-line phase shifters, the insertion loss  is caused by switch losses (which are independent of the phase-shift values) and line losses (which increase with the phase-shift value) \cite{microwaves101}. Nevertheless, the overall loss is dominated by the switch losses which implies that the phase-shifter loss is practically constant for different phase shifts. Therefore, in this paper, we assume that the phase-shifter efficiency is constant for different phase-shift values, i.e., $\rho_P(\beta)=\rho_P,\,\,\forall \beta$.}
\end{itemize}

{\BLUE
\begin{remk}
	In addition to \textit{RF losses}, each phase-shifter  \textit{consumes} a certain amount of \textit{power} in order to control its phase-shift states. The amount of power consumed depends on the adopted phase-shifter technology. For phase shifters realized by varactor diodes or micro-electro-mechanical system (MEMS) switches, the phase-shifter power consumption is almost negligible \cite{hum2013reconfigurable}. On the other hand, if the phase shifters are realized by positive-intrinsic-negative (PIN) diode switches, a constant direct current (DC) is needed to drive each switch which implies a constant power consumption per unit-cell element.  In this case, the total power consumption of the surface for controlling the phase shifters increases linearly with the number of  unit cells. Since we are interested in scalable MIMO transmitter architectures, we focus on the former realizations of the phase-shifters and neglect the corresponding power~consumption.
	\end{remk}
}

 Recall that the impact of the spillover, taper, and \textit{\BLUE antenna aperture} losses is included in  matrix $\mathbf{T}$ in (\ref{Eq:Tmaxtrix}). The remaining losses are accounted for in the power efficiency of IRS and ITS defined as {\BLUE $\rho_{\mathrm{srf}}=\rho^2_{P}\rho^{\rm srf}_{A}$ and $\rho_{\mathrm{srf}}=\rho_{P}\rho^{\rm srf}_{A}$}, respectively, which accounts for the combined effects of the surface aperture and phase-shifter losses.  In summary, the total power consumption of the IRS/ITS-aided MIMO architectures is obtained as
\begin{IEEEeqnarray}{lll}\label{Eq:PowerRA_TA}
	\Ptot &=\Pbb+N\Prfc+\frac{\Prd}{\rho_{\mathrm{pa}}}.
\end{IEEEeqnarray}
{\BLUE Due to the aforementioned power losses\footnote{Note that taper loss reduces the achievable rate but does not constitute a power loss.}, i.e., $\rho_{S}$,  $\rho_{P}$, and $\rho_{A}$, the active antennas have to transmit with power $\Prd$ to ensure the required power $\Ptx$ is radiated by the passive antennas where $\Prd\geq\Ptx$ holds. The following lemma relates the  power radiated by the active antennas, $\Prd$, to the power radiated by the intelligent surface $\Ptx$, based on matrix $\T$ and power efficiency factors $\rho_{S}$,  $\rho_{P}$, and $\rho_{A}$. For future reference, let $\varpi(\mathbf{T})=\frac{\sigma_{\max}(\mathbf{T})}{\sigma_{\min}(\mathbf{T})}\geq 1$ denote the condition number of matrix $\mathbf{T}$, where $\sigma_{\max}(\mathbf{T})$ and $\sigma_{\min}(\mathbf{T})$ denote the maximum and minimum singular values of $\mathbf{T}$, respectively. Moreover, although the values of  $\rho_{S}$ and $\rho_{A}$ may in general be different for each active antenna and depend on the position of the active antenna with respect to the intelligent surface, for simplicity of presentation, we assume identical $\rho_{S}$ and $\rho_{A}$ for all active antennas.}

{\BLUE
	\begin{lem} \label{Lem:PrdPtx}
		The  power radiated by the active antennas,  $\Prd$,  is bounded in terms of the transmit power radiated by the intelligent surface,  $\Ptx$, as  follows
	\begin{IEEEeqnarray}{lll}\label{Eq:PowerPtxPrd}
		\frac{\Ptx}{\varpi^2(\mathbf{T})\rho_{\rm rts}} &\leq\frac{\Ptx}{\sigma^2_{\max}(\T)} \nonumber\\
		&\leq\Prd \leq \frac{\Ptx}{\sigma^2_{\min}(\T)}\leq\frac{\varpi^2(\mathbf{T})\Ptx}{\rho_{\rm rts}},
	\end{IEEEeqnarray}
	where $\rho_{\rm rts}=\rho_{\rm srf}\rho_A^{\rm ant}\rho_S$. Therefore, assuming an ideal well-conditioned matrix $\T$ with  $\varpi(\mathbf{T})= 1$, we obtain $\Prd=\frac{\Ptx}{\rho_{\rm rts}}$. 
\end{lem}
	\begin{IEEEproof}
		The proof is provided in Appendix~\ref{App:LemPrdPtx}.
\end{IEEEproof}} 
 
 From \eqref{Eq:PowerRA_TA} and Lemma~\ref{Lem:PrdPtx}, we can conclude that  the total power consumption of IRS/ITS-aided antennas does not explicitly depend on the number of passive elements $M$ (or equivalently the size of the IRS/ITS) which makes them energy efficient and scalable.  Nevertheless, the condition number $\varpi(\mathbf{T})$  and the values of the spillover efficiency, $\rho_{S}$ are determined by factors such as the size of the intelligent surface, the beam pattern of the active antennas, the distance between the active antennas and the intelligent surface, etc., which may in turn be \textit{influenced} by $M$. Moreover, there is a trade-off between the spillover loss $\rho_{S}$ and the taper loss $\rho_{T}$ such that the former can be decreased at the expense of increasing the latter by employing a narrower beam for the active antennas \cite[Chapter~15]{balanis1982antenna}. 
  In the following subsection, we analytically show for a special case that  $\rho_{S}$ and $\rho_T$ can be made independent of $M$ by proper positioning of the antennas. Moreover, in Section~\ref{Sec:Sim}, we show via simulations that $\varpi(\mathbf{T})$ can be made small (i.e., close to one) when the antennas are properly~positioned.

\subsection{Special Case}\label{Sec:SpecialCase}
To illustrate the variation of the spillover and taper losses as a function of the feed pattern
and the angular extent of the intelligent surface, we consider the following simple class of axisymmetric feed antenna patterns which have been widely-adopted by the antenna community\footnote{\BLUE We note that our modeling and derivations are valid for general feed antenna patterns and only for the analysis in Section~\ref{Sec:SpecialCase} and the simulation results in Section~\ref{Sec:Sim}, we adopt the example  antenna pattern in \eqref{Eq:FeedPattern}.}  \cite{balanis1982antenna,pozar1997design}
\begin{IEEEeqnarray}{lll}\label{Eq:FeedPattern} 
G(\theta,\phi) \nonumber\\
= \begin{cases}
2(1+\kappa)\cos^\kappa(\theta),  & \mathrm{if}\,\,0\leq \theta \leq \frac{\pi}{2} \,\,\text{and}\,\, 0\leq\phi\leq 2\pi\\
0,\quad & \mathrm{otherwise}, \\
\end{cases}
\end{IEEEeqnarray}
where $\kappa\geq 2$ is a real number and normalization factor $2(1+\kappa)$ ensures that $\int_{\Omega} \frac{1}{4\pi} G(\phi,\theta)\mathrm{d}\Omega=1$ holds  \cite{balanis1982antenna}. Therefore, the (maximum) antenna gain in dB is $10\log_{10}(2(1+\kappa))$. The value of $\kappa$ (i.e., the gain of the active antenna) has to be jointly optimized with the relative position and orientation of the active antenna with respect to the intelligent surface   such that the active antenna illuminates only the intended part of the intelligent surface, see illumination strategies in Section~\ref{Sec:Illumination}. For the antenna pattern in \eqref{Eq:FeedPattern}  and assuming a \textit{circular} planar surface where the feed antenna orthogonally illuminates the center of the surface, the spillover loss is obtained as \cite[Chapter~15]{balanis1982antenna}
\begin{IEEEeqnarray}{lll} \label{Eq:RhoS}
\rho_{S}=\frac{\int_{\phi}\int_{\theta=0}^{\theta_0} G(\phi,\theta)\mathrm{d}\Omega}{\int_{\phi}\int_{\theta=0}^{\pi} G(\phi,\theta)\mathrm{d}\Omega}=1-\cos^{\kappa+1}(\theta_0),
\end{IEEEeqnarray}
where $\theta_0$ is the elevation angular extent of the intelligent surface with respect to the feed antenna. Similarly, the taper loss is obtained~as \cite[Chapter~15]{balanis1982antenna}
\begin{IEEEeqnarray}{lll} \label{Eq:RhoT}
	\rho_{T}&=\frac{\left[\int_{\phi}\int_{\theta=0}^{\theta_0} \sqrt{G(\phi,\theta)}\mathrm{d}S\right]^2}{2\pi (\cos^{-1}(\theta_0)-1) \int_{\phi}\int_{\theta=0}^{\pi} G(\phi,\theta)\mathrm{d}S}
	\nonumber\\
	&= \frac{\kappa-1}{\left(\kappa/2-1\right)^2}\times\frac{\left[1-\cos^{\kappa/2-1}(\theta_0)\right]^2}{(1-\cos^{\kappa-1}(\theta_0))(\cos^{-1}(\theta_0)-1)},\quad
\end{IEEEeqnarray}
where $\mathrm{d}S=\mathrm{d}S_{\theta}\mathrm{d}S_{\phi}$ is the normalized unit area covered by $[\theta, \theta+\mathrm{d}\theta]$ and $[\phi, \phi+\mathrm{d}\phi]$ on the intelligent surface,  $\mathrm{d}S_{\theta}=\cos^{-2}(\theta)\mathrm{d}\theta$, and $\mathrm{d}S_{\phi}=\sin(\theta)\mathrm{d}\phi$.  Moreover, the normalization factor $2\pi (\cos^{-1}(\theta_0)-1)$ ensures that $\rho_T=1$ for uniform illumination. Note that choosing a larger $\kappa$ decreases the spillover loss; however, it increases the taper loss.

The spillover and taper efficiencies in \eqref{Eq:RhoS} and \eqref{Eq:RhoT}, respectively, were derived for a circular planar surface. For square planar surfaces, we can obtain approximate expressions for $\rho_S$ and $\rho_T$ from \eqref{Eq:RhoS} and \eqref{Eq:RhoT}, respectively, by approximating the square surface with a circular surface having the same area. In particular, for a square surface with area $(\sqrt{M}d)^2$, the elevation angular extent  $\theta_0$ of the approximately equivalent circular surface is obtained as
	\begin{IEEEeqnarray}{lll}\label{Eq:Rd} 
		\theta_0\approx \tan^{-1}\left(\frac{d}{R_d}\sqrt{\frac{M}{\pi}}\right),
	\end{IEEEeqnarray}
where $R_d$ is the distance between the active antenna and the intelligent surface. Therefore, for square surfaces, if $R_d$ is chosen to be  proportional to $\sqrt{M}$, the value of $\theta_0$ is independent of $M$. Hence, in this case, the spillover and taper losses do not scale with $M$, cf. \eqref{Eq:RhoS} and \eqref{Eq:RhoT}.

\section{Precoder Design for IRS/ITS-Aided MIMO Architectures}\label{Sec:Precoder}

In this section, we propose two linear precoders for IRS/ITS-aided antennas exploiting the sparsity of the mmWave channel. We assume that CSI $\mathbf{H}$ is available at the transmitter and is used for precoder design. Therefore, similar to the precoder designs in \cite{el2014spatially,molisch17,ghanaatian2018feedback,alkhateeb2016frequency} for conventional MIMO architectures,  the frequency with which the proposed precoders (including the phase shifters at the intelligent surface) are updated should be chosen in accordance with the channel coherence time. This is in contrast to load modulated arrays  \cite{sedaghat2016load},  media-based modulation \cite{khandani2013media}, and  symbol-level precoding for IRS/ITS-aided MIMO \cite{bereyhi2019papr,bereyhi2020single}, where antenna loads or phase shifters change at the symbol rate. Exploiting the CSI,  ideally, we would like to design the optimal precoder {\BLUE which maximizes the achievable rate or equivalently the MI between $\mathbf{s}$ and $\mathbf{y}$, i.e.,  $R(\mathbf{F})\triangleq I(\mathbf{s};\mathbf{y})=\log_2\big|\mathbf{I}_{J}+\gamma\mathbf{H}\mathbf{F}\mathbf{F}\Herm\mathbf{H}\Herm\big|$,  for $\mathbf{y}$ given in \eqref{Eq:AWGN} and Gaussian $\mathbf{s}$}, as follows
\begin{IEEEeqnarray}{cll}\label{Eq:RateMax} 
	\underset{\mathbf{F}\in\mathcal{F}}{\mathrm{maximize}} \,\,&\log_2\big|\mathbf{I}_{J}+\gamma\mathbf{H}\mathbf{F}\mathbf{F}\Herm\mathbf{H}\Herm\big| 
	\nonumber\\
	 \mathrm{subject\,\,to:}\quad
	&\|\mathbf{F}\|_F^2\leq 1,
\end{IEEEeqnarray}
where $\gamma=\frac{\Ptx}{\sigma^2}$, constraint $\|\mathbf{F}\|_F^2\leq 1$ enforces the transmit power constraint, $\mathcal{F}=\{\mathbf{F}=\mathbf{DTB}|\mathbf{B}\in\mathbb{C}^{N\times Q} \,\,\text{and}\,\, \mathbf{D}=\Diag(d_1,\dots,d_M),$ $d_m\in\mathbb{A}\}$ is the set of feasible precoders, and $\mathbb{A}=\big\{x|x\in\mathbb{C} \,\,\text{and}\,\,|x|=1\big\}$ is the set of unit-modulus numbers. We note that the problem in \eqref{Eq:RateMax} is different from those considered in \cite{zang2019optimal,gao2016energy,yu2016alternating,el2014spatially,castanheira2017hybrid,ribeiro2018energy} for conventional MIMO architectures due to the different constraints imposed on the precoder via $\mathcal{F}$. Hence, the solutions proposed in the literature for conventional MIMO architectures are not directly applicable to  \eqref{Eq:RateMax}.

Unfortunately, problem (\ref{Eq:RateMax}) is not tractable since set $\mathcal{F}$ is not convex due to the unit-modulus constraint on the elements of phase-shift matrix $\mathbf{D}$. A similar challenge exists for conventional hybrid precoders where the elements of the corresponding analog precoder have to be unit-modulus. Since the global optimal solution of such non-convex problems cannot be found in a computationally efficient  manner,  suboptimal solutions have been pursued in the literature. These solutions can be classified into two categories: \textit{i)} solutions that find a local optimal/stationary point of the original problem \cite{zang2019optimal,gao2016energy,yu2016alternating}, and \textit{ii)} greedy solutions which exploit certain useful properties of the original problem (e.g., the sparsity of the mmWave channel) \cite{el2014spatially,castanheira2017hybrid,ribeiro2018energy}. In this section, we propose new precoder designs for IRS/ITS-aided MIMO which in part belong to both aforementioned categories. In particular, in each iteration, for a given phase-shift matrix, the digital baseband precoder is found as the global optimum of a corresponding sub-problem; however, for the phase-shift matrix, we only allow transmission in the direction of the AoDs of the channel which is an intuitive but in general heuristic choice. We note that IRS-aided antennas have been considered previously in \cite{zhou2018hardware} and a local optimal solution based on approximating the optimal FD precoder, instead of maximizing the MI as in (\ref{Eq:RateMax}), was derived.  We show in Section~\ref{Sec:Sim} via simulations that the proposed precoders outperform the benchmark scheme from \cite{zhou2018hardware}. 

\subsection{Rationale Behind the Proposed Precoders} 

For the spatially sparse channel model introduced in (\ref{Eq:Channel}),  $\mathcal{H}_t=\big\{\mathbf{h}_t(\theta_l^t,\phi_l^t),\forall l=1,\dots,L\big\}$ forms a vector space  for the rows of $\mathbf{H}$.  Moreover, in practice, $(\theta_l^t,\phi_l^t)$ is an RV that takes its values from a continuous distribution. Therefore, since $L\ll M$,  the elements of $\mathcal{H}_t$ are with probability one linearly independent  \cite{el2014spatially}. Let $\mathcal{H}_t^{\bot}$ denote the null space of $\mathcal{H}_t$. Thereby, any precoder $\mathbf{F}=\mathbf{F}_{\mathcal{H}_t}+\mathbf{F}_{\mathcal{H}_t^{\bot}}$ can be decomposed into matrix $\mathbf{F}_{\mathcal{H}_t}$ belonging to space $\mathcal{H}_t$ and matrix $\mathbf{F}_{\mathcal{H}_t^{\bot}}$ belonging to space $\mathcal{H}_t^{\bot}$. The following lemma formally characterizes the impact of $\mathbf{F}_{\mathcal{H}_t}$ and $\mathbf{F}_{\mathcal{H}_t^{\bot}}$ on the cost function and the constraint in (\ref{Eq:RateMax}).

\begin{lem}\label{Lem:ArraySpace}
For any given precoder $\mathbf{F}$, the relations $R(\mathbf{F})=R(\mathbf{F}_{\mathcal{H}_t})$ and $\|\mathbf{F}\|_F\geq\|\mathbf{F}_{\mathcal{H}_t}\|_F$ hold.  	
\end{lem} 
\begin{IEEEproof}
		The proof is given in Appendix~\ref{App:Lem_ArraySpace}. 	
\end{IEEEproof}

Motivated by the above results, we limit our attention to precoders of the form $\mathbf{F}=\mathbf{F}_{\mathcal{H}_t}$ which includes the optimal FD precoder \cite{ghanaatian2018feedback,el2014spatially}. More explicitly, the precoder is rewritten as $\mathbf{F}=\mathbf{H}_t\mathbf{C}$, where $\mathbf{H}_t=[\mathbf{h}_t(\theta_1^t,\phi_1^t),\dots,\mathbf{h}_t(\theta_L^t,\phi_L^t)]\in\mathbb{C}^{M\times L}$ and $\mathbf{C}\in\mathbb{C}^{L\times Q}$ contains the corresponding coefficients. For the FC hybrid MIMO architecture, the similarity of the structure of the optimal precoder $\mathbf{F}=\mathbf{H}_t\mathbf{C}$ and the corresponding hybrid precoder  $\mathbf{F}=\mathbf{R}\mathbf{B}$, where $\mathbf{B}$ and $\mathbf{R}$ denote the digital and analog precoders, receptively, has motivated researchers to use the channel response vectors $\mathbf{h}_t(\theta_l^t,\phi_l^t)$ as the columns of $\mathbf{R}$ \cite{gao2018low,el2014spatially,delgado2018feasible}. Since $\mathbf{R}$ has $N$ columns (i.e., there are $N$ RF chains), the hybrid precoder problem simplifies to choosing the best $N$ columns of $\mathbf{H}_t$ and the corresponding coefficients. Unfortunately, this concept is not directly applicable to the IRS/ITS precoder in (\ref{Eq:Precoder}) because of its different structure. Hence, we rewrite  $\mathbf{F}=\mathbf{H}_t\mathbf{C}$ in a more useful form. Let us divide the index set of the passive antennas $\{1,\dots,M\}$ into $N$ mutually exclusive sets $\mathcal{M}_{n},\,\,n=1,\dots,N$. {\BLUE We note that depending on the adopted illumination scenario, the passive antenna elements corresponding to set $\mathcal{M}_{n}$ may also receive signals from active antennas $n'\neq n$.} Now, $\mathbf{F}=\mathbf{H}_t\mathbf{C}$ can be rewritten~as
\begin{IEEEeqnarray}{lll} \label{Eq:OptimalFreform}
	\mathbf{F}=\sum_{n=1}^{N} \mathbf{H}_t^{\mathcal{M}_{n}}\mathbf{C},
\end{IEEEeqnarray}
where $\mathbf{H}_t^{\mathcal{M}_{n}}=\mathbf{I}_{\mathcal{M}_{n}}\mathbf{H}_t\in\mathbb{C}^{M\times L}$ and $\mathbf{I}_{\mathcal{M}_{n}}\in\{0,1\}^{M\times M}$ is a diagonal matrix whose $m$-th diagonal entry is one if $m\in\mathcal{M}_n$ and zero otherwise. In other words, we decompose $\mathcal{H}_t$ into $N$ subspaces, denoted by $\mathcal{H}_t^{\mathcal{M}_{n}},\,\,n=1,\dots,N$, which have mutually exclusive non-zero supports and are fully characterized by $\mathbf{H}_t^{\mathcal{M}_{n}},\,\,n=1,\dots,N$, respectively.  In a similar manner, let us rewrite the precoder in (\ref{Eq:Precoder}) as
\begin{IEEEeqnarray}{lll} \label{Eq:Freform}
	\mathbf{F}=\sum_{n=1}^{N} \mathbf{D}^{\mathcal{M}_{n}}\mathbf{T}^{\mathcal{M}_{n}}\mathbf{B},
\end{IEEEeqnarray}
where $\mathbf{D}^{\mathcal{M}_{n}}=\mathbf{I}_{\mathcal{M}_{n}}\mathbf{D}\in\mathbb{A}^{M\times N}$ and $\mathbf{T}^{\mathcal{M}_{n}}=\mathbf{I}_{\mathcal{M}_{n}}\mathbf{T}\in\mathbb{C}^{M\times N}$. Comparing (\ref{Eq:OptimalFreform}) and (\ref{Eq:Freform}) motivates us to choose $\mathbf{D}^{\mathcal{M}_{n}}$ such that $\mathbf{D}^{\mathcal{M}_{n}}\mathbf{T}^{\mathcal{M}_{n}}$ becomes similar to $\mathbf{H}_t^{\mathcal{M}_{n}}$. To do this, we have to address the following two challenges. First, since $\mathbf{D}^{\mathcal{M}_{n}}$ has only $M/N$ non-zero elements and $\mathbf{H}_t^{\mathcal{M}_{n}}$ has $ML/N$ non-zero elements, $\mathbf{H}_t^{\mathcal{M}_{n}}$ cannot be fully reconstructed via $\mathbf{D}^{\mathcal{M}_{n}}\mathbf{T}^{\mathcal{M}_{n}}$ as matrix $\mathbf{T}^{\mathcal{M}_{n}}$ is fixed. Hereby, we choose to reconstruct only one column of $\mathbf{H}_t^{\mathcal{M}_{n}}$ via $\mathbf{D}^{\mathcal{M}_{n}}\mathbf{T}^{\mathcal{M}_{n}}$. The unmatched columns of $\mathbf{D}^{\mathcal{M}_{n}}\mathbf{T}^{\mathcal{M}_{n}}$ are treated as interference. Fortunately, for $M\gg N$, the interference approaches zero due to channel hardening. Second, we have to choose which column of $\mathbf{H}_t^{\mathcal{M}_{n}}$ to reconstruct. In the following, we introduce two approaches, namely MI-based and OMP-based strategies, to choose the best $N$ columns of $\mathbf{H}_t^{\mathcal{M}_{n}}$. Note that the above precoder design effectively reduces the search space for the diagonal phase-shift matrix $\mathbf{D}$ from multi-dimensional continuous set $\mathbb{A}^{M\times 1}$ to the finite elements  of sets $\mathcal{H}_t^{\mathcal{M}_{n}},\,\,n=1,\dots,N$, i.e., $NL$ elements in total. Therefore, as typically $N,L\ll M$, one can adopt an exhaustive search over this reduced space to obtain the optimal (e.g., rate-maximizing) phase-shift matrix $\mathbf{D}$ for a given baseband precoder $\mathbf{B}$.

\subsection{MI-based Precoder}

For the proposed MI-based precoder, we design the phase-shift matrices $\mathbf{D}^{\mathcal{M}_{n}},\,\,n=1,\dots,N$, and the corresponding baseband precoder in an iterative manner, such that the MI expression in (\ref{Eq:RateMax}) is maximized \cite{alkhateeb2016frequency}. In particular, the proposed precoder design consists of an inner loop and an outer loop. The outer loop involves $N$ iterations where in the $n$-th iteration, we choose the best channel path for the design of phase-shift matrix $\mathbf{D}^{\mathcal{M}_{n}}$, and in the inner loop, we determine the corresponding baseband precoder, denoted by $\mathbf{B}_n$. In particular, the inner loop involves $L$ iterations where in the $l$-th iteration, we maximize the achievable rate by optimizing the baseband precoder $\mathbf{B}$ assuming $\mathbf{D}^{\mathcal{M}_{n}}$ is one of the elements of $\mathcal{H}_t^{\mathcal{M}_{n}}$. Therefore, we have to consider the following two problems:

\textbf{Optimizing the Baseband Precoder:} Here, we assume the phase-shift matrix $\mathbf{D}=\sum_{n=1}^{N} \mathbf{D}^{\mathcal{M}_{n}}$ is given. {\BLUE Since the logarithm is a monotonically increasing function}, the optimization problem for finding the digital baseband precoder $\mathbf{B}$, {\BLUE which maximizes the MI $I(\mathbf{s};\mathbf{y})$},  simplifies to
\begin{IEEEeqnarray}{cll}\label{Eq:DigitalPrecoderRate}
	\underset{\mathbf{B}\in\mathbb{C}^{N\times Q}}{\mathrm{maximize}} \,\,&\left|\mathbf{I}_{J}+\gamma \mathbf{H}\mathbf{C}_1\mathbf{B}\mathbf{B}\Herm\mathbf{C}_1\Herm\mathbf{H}\Herm\right| 
	\nonumber\\
	 \mathrm{subject\,\,to:}\quad
	& \Tr\left(\mathbf{C}_1\mathbf{B}\mathbf{B}\Herm \mathbf{C}_1\Herm\right)\leq 1,
\end{IEEEeqnarray}
where $\mathbf{C}_1=\mathbf{D}\mathbf{T}\in\mathbb{C}^{M\times N}$. Let us define matrix $\widetilde{\mathbf{H}}=\mathbf{H}\mathbf{C}_1\big(\mathbf{C}_1\Herm\mathbf{C}_1\big)^{-\frac{1}{2}}\in\mathbb{C}^{J\times N}$ and its corresponding singular value decomposition (SVD) $\widetilde{\mathbf{H}}=\mathbf{U}\boldsymbol{\Sigma}\mathbf{V}\Herm$, where $\mathbf{U}=[\mathbf{u}_1,\dots,\mathbf{u}_J]\in\mathbb{C}^{J\times J}$ and $\mathbf{V}=[\mathbf{v}_1,\dots,\mathbf{v}_N]\in\mathbb{C}^{N\times N}$ are unitary matrices containing the left and right singular vectors, respectively, and $\boldsymbol{\Sigma}$ is a diagonal matrix containing the singular values $\sigma_1,\dots,\sigma_N$ in descending order. The solution of \eqref{Eq:DigitalPrecoderRate} is given in the following~lemma.

\begin{lem}\label{Lem:OptB}
For a given phase-shift matrix $\mathbf{D}$, the optimal baseband precoder $\mathbf{B}$ as a solution of \eqref{Eq:DigitalPrecoderRate}  is given~by
\begin{IEEEeqnarray}{rll}\label{Eq:BsolRate} 
	\mathbf{B} = \big(\mathbf{C}_1\Herm\mathbf{C}_1\big)^{-\frac{1}{2}}[\mathbf{v}_1,\dots,\mathbf{v}_Q]\mathbf{Z},
\end{IEEEeqnarray}
where $\mathbf{Z}=\Diag(\sqrt{z_1},\dots,\sqrt{z_Q})\in\mathbb{C}^{Q\times Q}$,  $z_q = \big[\mu-\frac{1}{\gamma\sigma_q^2}\big]^+$, and threshold $\mu$ is chosen such that constraint $\sum_{q=1}^Qz_q=1$ is~met. 
\end{lem}
\begin{IEEEproof}
		The proof is given in Appendix~\ref{App:LemOptB}.	
\end{IEEEproof}

Note that the baseband precoder $\mathbf{B}$ effectively eliminates the interference between the data streams.

\textbf{Optimizing the Phase-Shift Matrix:} As discussed earlier, we decompose $\mathbf{D}$ into $N$ components $\mathbf{D}^{\mathcal{M}_{n}},\,\,n=1,\dots,N$, which are initialized to the identity matrix $\mathbf{I}_{M}$ and their values are updated in each iteration. In particular, in the $n$-th iteration, the following problem is solved
\begin{IEEEeqnarray}{rll} \label{Eq:AnalogPrecoderRateRelaxed}
	\underset{\mathbf{D}^{\mathcal{M}_{n}}\in\mathcal{D}_n}{\mathrm{maximize}} \,\,&\Big|\mathbf{I}_{J}+\gamma \mathbf{H}\mathbf{D}\mathbf{C}_2\mathbf{D}\Herm\mathbf{H}\Herm\Big|,
\end{IEEEeqnarray}
where $\mathbf{C}_2=\mathbf{TB}(\mathbf{D})\mathbf{B}(\mathbf{D})\Herm\mathbf{T}\Herm\in\mathbb{C}^{M\times M}$. Here, $\mathbf{B}(\mathbf{D})$ denotes the optimal baseband precoder as a function of a given phase-shift matrix $\mathbf{D}$ which is obtained from (\ref{Eq:BsolRate}). Moreover, set $\mathcal{D}_n$ is given by 
\begin{IEEEeqnarray}{lll} \label{Eq:Dset}
	\mathcal{D}_n = \Big\{\mathbf{D}^{\mathcal{M}_{n}}[l]\in\mathbb{A}^{M\times M},\,\,\forall l=1,\dots,L\big| 
	\nonumber\\
	\mathbf{D}^{\mathcal{M}_{n}}_{m,m'}[l]  =  	
	\begin{cases}
		\exp\big(j\big[\angle (\mathbf{H}_t^{\mathcal{M}_{n}})_{m,l} 
		\!\!\!\!&- \angle \mathbf{T}_{m,n}\big]\big),\\
		&\forall m=m'\in\mathcal{M}_n \\
		0, &\mathrm{otherwise}
	\end{cases}\Big\}.\quad\,\,\,
\end{IEEEeqnarray}
As can be seen, the cardinality of $\mathcal{D}_n$ is $L$ which allows us to solve (\ref{Eq:AnalogPrecoderRateRelaxed}) via an exhaustive search. Algorithm~\ref{Alg:PrecoderMI} summarizes the main steps of the proposed MI-based precoder design.

\begin{algorithm}[t]
	\caption{MI-based Precoder Design}
	\begin{spacing}{1}
		\begin{algorithmic}[1]\label{Alg:PrecoderMI}
			\STATE \textbf{initialize:} $\mathbf{D}^{\mathcal{M}_{n}}=\mathbf{I}_{M},\,\,\forall n$.  
			\FOR{$n=1,\dots,N$}
			\FOR{$l=1,\dots,L$}
			\STATE Set $\mathbf{D}^{\mathcal{M}_{n}}=\mathbf{D}^{\mathcal{M}_{n}}[l]$ from $\mathcal{D}_n$ in (\ref{Eq:Dset}).
			\STATE Find $\mathbf{B}[l]$ using (\ref{Eq:BsolRate}) for  $\mathbf{D}[l]=\sum_{i=1}^n\mathbf{D}^{\mathcal{M}_{i}}$.
			\STATE Set $R[l]=\log_2\big(\big|\mathbf{I}_{J}+\gamma\mathbf{H}\mathbf{F}\mathbf{F}\Herm\mathbf{H}\Herm\big|\big)$ for $\mathbf{F}=\mathbf{D}[l]\mathbf{TB}[l]$.
			\ENDFOR
			\STATE Update $\mathbf{D}^{\mathcal{M}_{n}}=\mathbf{D}^{\mathcal{M}_{n}}[l^*]$ and $\mathbf{B}_n=\mathbf{B}[l^*]$ for $l^*=\underset{l}{\mathrm{argmax}}\,\,R[l]$.
			\ENDFOR
			\STATE Return $\mathbf{D}=\sum_{n=1}^{N} \mathbf{D}^{\mathcal{M}_{n}}$ and $\mathbf{B}=\mathbf{B}_N$.\vspace{0.05cm}
		\end{algorithmic}
	\end{spacing}
\end{algorithm}

\subsection{OMP-based Precoder}
In this subsection, we propose a second precoder, namely the OMP-based precoder, which is computationally less complex than the MI-based precoder (cf. Section~\ref{Sec:Complexity}) but achieves a similar performance in poor scattering environments  (cf. Section~\ref{Sec:Sim}). In particular, the OMP-based precoder attempts to approximate the optimal unconstrained precoder for the FD MIMO architecture, denoted by $\mathbf{F}^{\mathrm{opt}}$, using the OMP algorithm.   Minimization of $\|\mathbf{F}^{\mathrm{opt}}-\mathbf{F}\|_F$ has been commonly adopted  as design criterion for constrained hybrid precoders for conventional MIMO architectures, see e.g. \cite{el2014spatially,zhou2018hardware,lin2016energy}. Motivated by this, we consider the following optimization problem for the precoder of IRS/ITS-aided MIMO systems
\begin{IEEEeqnarray}{cll} \label{Eq:ProbMin}
	\underset{\mathbf{B}\in\mathbb{C}^{N\times Q},\mathbf{D}^{\mathcal{M}_{n}}\in\mathcal{D}_n}{\mathrm{minimize}} \,\, &\big\|\mathbf{F}^{\mathrm{opt}} -  \mathbf{D}\mathbf{T}\mathbf{B} \big\|_F^2 
	\nonumber\\
	\mathrm{subject\,\,to:} & \|\mathbf{DTB}\|_F^2\leq 1.
\end{IEEEeqnarray}
Again, let us fix sets $\mathcal{M}_{n},\,\,n=1,\dots,N$, a priori.  The proposed precoder employs $N$ iterations where in each iteration, the following two problems are solved:

\textbf{Optimizing the Phase-Shift Matrix:} Let $\mathbf{F}_n^{\mathrm{res}}=\mathbf{F}^{\mathrm{opt}}-\sum_{i=1}^{n} \mathbf{D}^{\mathcal{M}_{i}}\mathbf{T}^{\mathcal{M}_{i}}\mathbf{B}_n$ denote the residual precoder in iteration $n$ where $\mathbf{B}_{n}$ is the baseband precoder designed in iteration~$n$. In each iteration, we project the residual matrix from the previous iteration onto the space defined by $\mathbf{H}_t$ and find the direction $l^*$ that has the maximum projected value. This can be mathematically formulated as follows
\begin{IEEEeqnarray}{lll} \label{Eq:Angle}
l_n^*=\underset{l=1,\dots,L}{\mathrm{argmax}}\,\,(\boldsymbol{\Psi}\boldsymbol{\Psi}\Herm)_{l,l},
\end{IEEEeqnarray}
where $\boldsymbol{\Psi}=\mathbf{H}_t\Herm \mathbf{F}_{n-1}^{\mathrm{res}}\in\mathbb{C}^{L\times Q}$.   Therefore,  $\mathbf{D}^{\mathcal{M}_{n}}$ is selected as the element of  $\mathcal{D}_n$ corresponding to the $l^*$-th channel path, cf. \eqref{Eq:Channel}.   
 
\textbf{Optimizing the Baseband Precoder:} By defining $\mathbf{C}_3=\sum_{i=1}^{n} \mathbf{D}^{\mathcal{M}_{i}}\mathbf{T}^{\mathcal{M}_{i}}\in\mathbb{C}^{M\times N}$, we can formulate the  optimization problem for $\mathbf{B}_n$ as follows
\begin{IEEEeqnarray}{cll} \label{Eq:ProbB}
	\underset{\mathbf{B}\in\mathbb{C}^{N\times Q}}{\mathrm{minimize}}\,\, &\big\|\mathbf{F}^{\mathrm{opt}} -  \mathbf{C}_3\mathbf{B} \big\|_F^2,
	\nonumber\\
	 \mathrm{subject\,\,to:} \quad &\|\mathbf{C}_3\mathbf{B}\|_F^2\leq 1,
\end{IEEEeqnarray}
 which has the following well-known normalized least square solution \cite{el2014spatially} 
\begin{IEEEeqnarray}{lll} \label{Eq:SolB}
\mathbf{B}_n  = \frac{(\mathbf{C}_3\Herm\mathbf{C}_3)^{-1}\mathbf{C}_3\Herm\mathbf{F}^{\mathrm{opt}}}
{\|\mathbf{C}_3(\mathbf{C}_3\Herm\mathbf{C}_3)^{-1}\mathbf{C}_3\Herm\mathbf{F}^{\mathrm{opt}}\|_F}.
\end{IEEEeqnarray}
Algorithm~\ref{Alg:PrecoderOMP} summarizes the main steps for the proposed OMP-based precoder design.

\begin{algorithm}[t]
	\caption{OMP-based Precoder Design}
	\begin{spacing}{1}
		\begin{algorithmic}[1]\label{Alg:PrecoderOMP}
			\STATE \textbf{initialize:} $\mathbf{F}_0^{\mathrm{res}}=\mathbf{F}^{\mathrm{opt}}$ and $\mathbf{D}^{\mathcal{M}_{n}}=\boldsymbol{0}_{M\times M},\,\,\forall n$.  
			\FOR{$n=1,\dots,N$}
			\STATE $l_n^*=\mathrm{argmax}_{l=1,\dots,L}\,\,(\boldsymbol{\Psi}\boldsymbol{\Psi}\Herm)_{l,l}$ 
			for $\boldsymbol{\Psi}=\mathbf{H}_t\Herm \mathbf{F}_{n-1}^{\mathrm{res}}$.
			\STATE Update $\mathbf{D}^{\mathcal{M}_{n}}$ as the element of set $\mathcal{D}_n$ in (\ref{Eq:Dset}) corresponding to $l_n^*$.
			\STATE Update $\mathbf{B}_n$ using (\ref{Eq:SolB}) for $\mathbf{C}_3=\sum_{i=1}^{n} \mathbf{D}^{\mathcal{M}_{i}}\mathbf{T}^{\mathcal{M}_{i}}$.
			\STATE Update $\mathbf{F}_n^{\mathrm{res}}=\mathbf{F}^{\mathrm{opt}}-\sum_{i=1}^{n} \mathbf{D}^{\mathcal{M}_{i}}\mathbf{T}^{\mathcal{M}_{i}}\mathbf{B}_n$.
			\ENDFOR
			\STATE Return $\mathbf{D}=\sum_{n=1}^{N} \mathbf{D}^{\mathcal{M}_{n}}$ and $\mathbf{B}=\mathbf{B}_N$.\vspace{0.05cm}
		\end{algorithmic}
	\end{spacing}
\end{algorithm}

\subsection{Complexity Analysis}\label{Sec:Complexity}

Let us assume that $M\gg J\geq N \geq Q$ and $M\gg L\geq N \geq Q$ hold. Moreover, we use the following results: The SVD of matrix $\mathbf{A}\in\mathbb{C}^{m\times n}$ of rank $p$ has complexity order $O(mnp)$, the inversion of matrix $\mathbf{A}\in\mathbb{C}^{m\times m}$ has complexity order $O(m^3)$, and the multiplication $\mathbf{AB}$ of matrices $\mathbf{A}\in\mathbb{C}^{m\times n}$ and $\mathbf{B}\in\mathbb{C}^{n\times p}$ has complexity order $O(mnp)$ \cite{najafi2019cran,golub2012matrix}. The MI-based precoder involves $NL$ iterations (i.e., inner and outer loops) where each iteration comprises the SVD of matrix $\widetilde{\mathbf{H}}$ (i.e., $O(JN^2)$), the inversion of matrix $\mathbf{C}_1\Herm\mathbf{C}_1$ (i.e., $O(N^3)$), and matrix multiplications (i.e., $O(MNJ)$). Hence, recalling $M\gg J, N, Q$, the overall complexity order of the MI-based precoder is $O(MN^2JL)$.
On the other hand, the OMP-based precoder requires the SVD of matrix $\mathbf{H}$ (i.e., $O(MJL)$) and involves $N$ iterations where each iteration comprises the inversion of matrix $\mathbf{C}_3\Herm\mathbf{C}_3$ (i.e., $O(N^3)$) and matrix multiplications (i.e., $O(MLQ+MN^2)$). Assuming $N=Q$, the overall complexity order of the OMP-based precoder simplifies to $O(M(N^2+J)L)$.  {\BLUE In summary, the complexity of both proposed algorithms is linear in $M$, which is a crucial advantage for ultra massive MIMO transmitters employing several hundreds (or even thousands) of transmit antennas. Moreover, computing the OMP-based precoder entails a lower complexity than computing the MI-based precoder as $N^2+J<N^2J$ holds for typical values of $N$ and $J$, i.e., $N,J>1$.}

\section{Simulation Results}\label{Sec:Sim} 
In this section, we first describe the considered simulation setup and benchmark schemes. Subsequently, we study the performance of the proposed precoders and the impact of the system parameters. Then, we compare  the performance of IRS/ITS-aided MIMO with that of the conventional MIMO architectures.  Finally, we study the impact of imperfect CSI on the performance of the proposed precoder design.

\subsection{Simulation Setup}
We generate the channel matrices according to (\ref{Eq:Channel}). Thereby, we assume that the AoAs/AoDs $\theta_l^t$, $\theta_l^r$,  $\phi_l^t$, and $\phi_l^r$ are  uniformly distributed RVs in the intervals given in Table~I. Moreover, we use the square uniform planar array in (\ref{Eq:UPA}), i.e., a $\sqrt{M}d\times\sqrt{M}d$ planar surface. The channel coefficient for each effective path is modeled as $h_l=\sqrt{\bar{h}_l}\tilde{h}_l$, where $\bar{h}_l$ and $\tilde{h}_l$ are the path loss and the random fading components, respectively, and are given~by
\begin{IEEEeqnarray}{lll}\label{Eq:PathLossRayleigh}
	\bar{h}_l = \left(\frac{\lambda}{4\pi \ell}\right)^{\eta} 
	\quad \text{and} \quad
	\tilde{h}_l = \mathcal{CN}(0,1),
\end{IEEEeqnarray}
respectively. In (\ref{Eq:PathLossRayleigh}), $\ell$ denotes the distance between the transmitter and the receiver and $\eta$ represents the path-loss exponent. The noise power at the receiver is given by $\sigma^2=WN_0N_F$, where $W$ is the bandwidth, $N_0$ represents the noise power spectral density, and $N_F$ denotes the noise figure.  All results shown in this	section have been averaged over $10^3$ random realizations of 
	channel matrix $\mathbf{H}$ from (\ref{Eq:Channel}) which includes the random realizations of the AoAs $(\theta_l^r,\phi_l^r)$, the AoDs $(\theta_l^t,\phi_l^t)$, and the corresponding path coefficients $h_l$.

We arrange the active antennas with respect to the intelligent surface as shown in Fig.~\ref{Fig:Illumination} and described in Section~\ref{Sec:Illumination}. We neglect the impact of the blockage of the active antennas in IRS-aided antennas; nevertheless, we study the performance of IRS-aided antennas under blockage-free illumination which is designed to avoid the blockage of desired AoAs/AoDs \cite{arrebola2008multifed,pozar1997design}.  Moreover, we adopt the feed antenna pattern in (\ref{Eq:FeedPattern}), which is widely used in the antenna community  \cite{balanis1982antenna,pozar1997design}. {\BLUE For the passive antenna elements, we assume a uniform antenna pattern in the \textit{half-space} where the intelligent surface faces the active antennas, which implies a constant antenna gain of $3$~dB.}  Unless otherwise stated, we adopt the default values of the system parameters  provided in Table~I, which include the values of the parameters of the conventional MIMO architectures too, cf. Section~\ref{Sec:Benchmark}. {\BLUE  Recall that the analysis in Section~\ref{Sec:SpecialCase} revealed that $R_d$ should scale with the square of the area of the passive surface  that is intended to be illuminated.  Therefore, for FI and blockage-free PI,  the value of $R_d$ has to scale with $\sqrt{M}$ such that the elevation angular extent $\theta_0$ of the \textit{entire array} with respect to any of the active antennas remains approximately constant, cf. \eqref{Eq:Rd}.} Furthermore, we assume a fixed value for $R_r$ which is larger than $\lambda/2$ to ensure negligible mutual coupling among the active antennas, cf. Assumption A3. {\BLUE In contrast, for PI and SI, the value of $R_d$ has to scale with $\sqrt{M/N}$ such that the elevation angular extent of each \textit{sub-array} with respect to its dedicated active antenna remains approximately constant.} Moreover, we scale $R_r$ with $M$ such that each active antenna remains close to the center of the area where the passive elements it serves are located. Therefore, in Table~I, we provide the values of $R_r$ and $R_d$ for two scenarios, namely scenario S1 for FI and blockage-free PI, and scenario S2 for PI and SI.

\begin{table*}
	\label{Table:Parameter}
	\caption{Default Values of System Parameters \cite{yan2018performance,ribeiro2018energy,lin2016energy,lau2012reconfigurable,di2015reconfigurable,wang2017spectrum,akdeniz2014millimeter}.\vspace{-0.3cm}} 
	\begin{center}
		\scalebox{0.6}{
			\begin{tabular}{|| c | c | c | c | c | c | c | c | c | c | c | c | c | c || }
				\hline\vspace{-0.3cm}
				& & & & & &  & & & & & & &\\ 
				Parameter & $\ell$ & $\eta$ & $L$ & $\theta_l^t$, $\theta_l^r$ & $\phi_l^t$, $\phi_l^r$ & $N_0$ & $N_F$ & $W$ 
				& $\lambda$ & $d$ & $R_d$ & $R_r$ & $\kappa$  \\ \hline\hline
				\vspace{-0.35cm}
				& & & & & &  & & & & & & & \\
				Value & $100$~m  & $2$ & $8$ & $[-2\pi/3,2\pi/3]$ & $[-\pi/2,\pi/2]$ & $-174$~dBm/Hz & $6$ dB & $100$ MHz & $10$~mm ($28$GHz)  & $\lambda/2$ & S1:$\frac{4d\sqrt{M}}{\sqrt{\pi}}$, S2:$\frac{4d\sqrt{M}}{\sqrt{N\pi}}$ & S1:$2d$, S2:$\frac{d\sqrt{2M}}{4}$ 
				& $49$ (20~dB gain)   \\ \hline 
			\end{tabular}
		} 
		
		\vspace{0.3cm}
		
		\scalebox{0.6}{
			\begin{tabular}{|| c | c | c | c | c | c | c | c | c | c | c | c | c | c | c | c ||}
				\hline\vspace{-0.3cm}
				& & & & & & & & & & &  & & & &\\ 
				Parameter &  $\Pbb$ & $\Prfc$ & $\Psw$   & $\Ptx$ & $\Pamp$ & $G_{\mathrm{amp}}$ & $L_D$& $L_C$ & $L_P$ ($1/\rho_P$ in dB) & $1/\rho^{\rm srf}_A$ in dB & $\rho_{\mathrm{amp}}$ & $Q$ & $N$ & $M$ & $J$  \\ \hline\hline
				\vspace{-0.35cm}
				& & & & & & & & & & &  & & & &\\
				Value  &  $200$ mW & $100$ mW & $5$ mW  & $20$ dBm & $40$ mW & $10$ dB& $3.6$ dB & $3.6$ dB & $2$ dB & IRS: $0.5$ dB, ITS: $1.5$ dB & $0.3$ & $4$ & $4$ & $256$ & $16$ \\ \hline 
			\end{tabular}
		}\vspace{-0.5cm}
	\end{center}
\end{table*}

\subsection{Benchmark Schemes}\label{Sec:Benchmark}

We consider the FD, FC hybrid, PC hybrid, and LA antennas as benchmark architectures. The precoder structures and power consumption models for these architectures are summarized in Table II. In the following, we briefly explain the assumptions made to arrive at these expressions for conventional MIMO systems.

\textbf{Fully-Digital MIMO:} Here, we have $N=M$ RF chains which enable FD precoding, i.e., $\mathbf{F}=\mathbf{B}$, where $\mathbf{B}$ is the digital precoder. For FD MIMO, we adopt the optimal unconstrained precoder obtained from the SVD of the channel and water filling power allocation \cite{ghanaatian2018feedback}. 

\textbf{Fully-Connected Hybrid MIMO:} In the FC hybrid architecture, we have $N$ RF chains whose outputs are connected to $M$ antennas via passive analog dividers, phase shifters, and combiners \cite{el2014spatially}. For this MIMO architecture, the precoder is given by $\mathbf{F} = \mathbf{R}\mathbf{B}$, where $\mathbf{B}\in\mathbb{C}^{N\times Q}$ denotes the digital precoder and $\mathbf{R}\in\mathbb{A}^{M\times N}$ represents the analog RF precoder. We adopt the spatially-sparse precoder introduced in \cite{el2014spatially}. 
For large RF networks, the insertion loss may easily exceed $20$-$30$~dB which makes a one-shot power compensation infeasible due to amplifier nonlinearities at high gains \cite{yan2018performance}. In practice, to compensate for this insertion loss, multiple gain-compensation amplifiers (GCAs) are cascaded to ensure that a minimum power is delivered to drive the PAs before transmission via the antennas \cite{yan2018performance,Rodriguez2016_MIMO_Loss}.  Assuming that the signal is amplified by GCAs before being fed to the PAs to compensate for the RF losses\footnote{In practice, multiple stages of power amplification are needed within the RF network to ensure that the signal power does not get too weak, see \cite{yan2018performance} for examples of multiple-stage power amplification. Our motivation for considering single-stage power amplification in this paper is two-fold. First, the exact design of multiple-stage amplification crucially depends on the specific system parameters, e.g., $M,N,G_{\mathrm{amp}}$, and $L_{\mathrm{rf}}$, and cannot be easily generalized. Second, since the number of required GCAs is larger for multiple-stage amplification, single-stage amplification constitutes a favorable choice for hybrid MIMO architectures, which we consider as performance benchmarks.}, the power consumption of the RF network is given by $\lceil\frac{L_{\mathrm{rf}}}{G_{\mathrm{amp}}}\rceil \Pamp$, where $L_{\mathrm{rf}}$ is the total loss in dB occurring in the RF network, $G_{\mathrm{amp}}$ denotes the maximum amplification gain of the GCAs in dB, and $\Pamp$ is their respective power consumption. Assuming that the power dividers (combiners) are implemented by a cascade of two-port power dividers (combiners), we need at least $\lceil\log_2(M)\rceil$ ($\lceil\log_2(N)\rceil$) stages of division (combining) \cite{Rodriguez2016_MIMO_Loss,Pozar2009microwave}. Therefore, the total power loss for the signal flowing towards each antenna is obtained as $L_{\mathrm{rf}}=\lceil\log_2(M)\rceil L_D+\lceil\log_2(N)\rceil L_C+L_P$, where $L_D$ ($L_C$) is the power loss of each three-port divider (combiner) in dB and $L_P=10\log_{10}(1/\rho_P)$ \cite{Rodriguez2016_MIMO_Loss,Pozar2009microwave}.  

\textbf{Partially-Connected MIMO:}  The signal model for the PC architecture is identical to that of the FC architecture, i.e., $\mathbf{F} = \mathbf{R}\mathbf{B}$, with the difference that $\mathbf{R}$ is now a block-diagonal matrix $\mathbf{R}=\Diag(\mathbf{r}_1,\dots,\mathbf{r}_N), \mathbf{r}_n\in\mathbb{A}^{r_n\times 1}$, where $\mathbf{r}_n\in\mathbb{A}^{r_n\times 1}$ is the RF precoder vector which connects the output of the $n$-th RF chain to $r_n$ antennas \cite{yu2016alternating,gao2016energy}. Note that $\sum_{n=1}^N r_n=M$ has to hold. We assume that all RF chains are connected to the same number of antennas, i.e., $r_n=M/N,\,\,\forall n$, where we assume that $N$ is a divisor of $M$. Equivalently, the precoder for the PC architecture can be rewritten as $\mathbf{F}=\mathbf{D}\tilde{\mathbf{T}}\mathbf{B}$ where $\tilde{\mathbf{T}}$ is a fixed matrix whose element in the $m$-th row and $n$-th column is one if the $m$-th antenna is connected to the $n$-th RF chain and zero otherwise and diagonal matrix $\mathbf{D}$ is the corresponding analog precoder. {\BLUE Considering the structure of $\mathbf{F}$, the proposed Algorithms~1 and 2 can be exploited for precoder design for the PC hybrid MIMO architecture.  In Fig.~\ref{Fig:RatePowerEE_M}, where we compare PC MIMO and IRS/ITS-aided MIMO, we employ Algorithm~2 for both architectures to ensure a fair comparison.} Noting that the PC architecture does not include a power combiner and assuming that the power dividers are implemented by a cascade of two-port power dividers, we obtain the total power consumption given in Table~II.

\begin{table*}
	\label{Table:Comparison}
	\caption{Comparison of Different MIMO Architectures, namely FD, FC, PC, LA, IRS-aided, and ITS-aided antennas.  } 
	\begin{center}
		\vspace{-0.1cm}
		\scalebox{0.6}{
			\begin{tabular}{|| c | c | c | c ||}
				\hline 
				Architecture & Precoder $\mathbf{F}$ & Constraints & Total Power Consumption $\Ptot$ \\ \hline \hline 
				\vspace{-0.4cm} & & & \\  
				FD & $\mathbf{B}$ & $\mathbf{B}\in\mathbb{C}^{M\times Q}$ & $\Pbb+M\Prfc+\frac{\Ptx}{\rho_{\mathrm{pa}}}$\\ \hline
				\vspace{-0.4cm} & & & \\   
				FC    &  $\mathbf{R} \mathbf{B}$ & $\mathbf{B}\in\mathbb{C}^{N\times Q}$, $\mathbf{R}\in\mathbb{A}^{M\times N}$ & $\Pbb+N\Prfc+\big\lceil \frac{\lceil\log_2(M)\rceil L_D+\lceil\log_2(N)\rceil L_C+L_P}{G_{\mathrm{amp}}}\big\rceil M\Pamp+\frac{\Ptx}{\rho_{\mathrm{pa}}}$\\ \hline  
				\vspace{-0.4cm} & & & \\ 
				PC    &  $\mathbf{R} \mathbf{B}$ & $\mathbf{B}\in\mathbb{C}^{N\times Q}$, $\mathbf{R}=\Diag(\mathbf{r}_1,\dots,\mathbf{r}_N), \mathbf{r}_n\in\mathbb{A}^{r_n\times 1}$, $\sum_{n=1}^N r_n=M$ & $\Pbb+N\Prfc+\Big\lceil \frac{\lceil\log_2(M/N)\rceil L_D+L_P}{G_{\mathrm{amp}}}\Big\rceil M\Pamp+\frac{\Ptx}{\rho_{\mathrm{pa}}}$\\ \hline   
				\vspace{-0.4cm} & & & \\
				LA   &  $\mathbf{DTSB}$ & $\mathbf{B}\in\mathbb{C}^{N\times Q}$, $\mathbf{S}\in\{0,1\}^{K\times N}$, fixed matrix $\mathbf{T}\in\mathbb{C}^{M\times N}$, fixed $\mathbf{D}=\Diag(d_1,\dots,d_M), d_m\in\mathbb{A}$ &$\Pbb+N\Prfc+N\Psw+\frac{\Ptx}{\rho_{\mathrm{rts}}\rho_{\mathrm{pa}}}$\\ \hline   
				IRS \& ITS    &  $\mathbf{DTB}$ & $\mathbf{B}\in\mathbb{C}^{N\times Q}$, fixed matrix $\mathbf{T}\in\mathbb{C}^{M\times N}$ (cf. (\ref{Eq:Tmaxtrix})), $\mathbf{D}=\Diag(d_1,\dots,d_M), d_m\in\mathbb{A}$ &$\Pbb+N\Prfc+\frac{\Ptx}{\rho_{\mathrm{rts}}\rho_{\mathrm{pa}}}$\\ \hline  
			\end{tabular}
		} 
	\end{center}\vspace{-0.3cm}
\end{table*}

\textbf{Lens Array MIMO:} EM lenses are phase-shifting devices which can be designed employing either an array of passive antenna elements (similar to the considered ITS)  \cite{popovic2002multibeam} or  continuous aperture phase shifting \cite{gao2018low,brady2013beamspace}. We consider the former option since it allows us to employ similar surfaces for LA and ITS-aided antennas. Therefore, the precoder for LA antennas is given by $\mathbf{F} = \mathbf{D}\mathbf{T}\mathbf{S}\mathbf{B}$, 
where $\mathbf{B}\in\mathbb{C}^{N\times Q}$ is the digital precoder matrix, $\mathbf{S}\in\{0,1\}^{K\times N}$ is a binary switching matrix which specifies which RF chain is connected to which active antenna, $\mathbf{T}\in\mathbb{C}^{M\times K}$ is a fixed matrix which models the channel between the active and the passive surface (similar as for IRS/ITS-aided antennas  in (\ref{Eq:Tmaxtrix})), and $\mathbf{D}\in\mathbb{A}^{M\times M}$ is a fixed diagonal matrix with unit-modulus elements designed with the objective to focus the wavefront perpendicular to the lens plane at the focal point of the lens (i.e., a non-reconfigurable/non-intelligent surface). For LA antennas, we use a modification of the proposed OMP-based precoder obtained with Algorithm~2. Since for LA antennas $\mathbf{D}$ is fixed and different active antennas are selected via matrix $\mathbf{S}$ for transmission in different directions, the main change required when adapting Algorithm~2 to LA antennas is that line~4 is replaced with the selection of the active antenna which is used to transmit the signal of RF chain $n$ along the AoD $(\theta_{l_n^*}^t,\phi_{l_n^*}^t)$ chosen in line~3. Here, we adopt the following antenna selection strategy: 
\begin{IEEEeqnarray}{lll}\label{Eq:AntennaSelection}
	[\mathbf{S}]_{k,n} = \begin{cases}
		1,&\mathrm{if}\,\,k=\underset{\tilde{k}=1,\dots,K}{\mathrm{argmax}}\,\,\left|(\mathbf{h}_t\Herm(\theta_{l_n^*}^t,\phi_{l_n^*}^t)\mathbf{DT})_{\tilde{k}}\right|\\
		0,&\mathrm{otherwise.}
	\end{cases}
\end{IEEEeqnarray}
Moreover, we assume that in the LA architecture, the active antennas are placed on the focal arc of the lens as given in \cite[Eqs.~(1) and (2)]{zeng2018multi} with focal distance $4\lambda\sqrt{M/\pi}$, and illuminate the center of the passive EM lens. For the design in \cite{zeng2018multi}, the number of active antennas is a linear function of the effective lens aperture or equivalently $M$, which we refer to as LA antennas with full $K$. In addition, we consider the case where $K$ is fixed independent of $M$ and refer to it as LA antennas with fixed $K$. The power consumption model for LA antennas is similar to that for ITS-aided antennas except for the additional power consumed by the $N$ active switches, i.e., $N\Psw$, where $\Psw$ denotes the power consumption of each switch~\cite{wang2017spectrum}.

\subsection{Performance of Proposed Precoders and Impact of the System Parameters}

\begin{figure}\vspace{-0.5cm}
	\centering 
	\includegraphics[width=1\linewidth]{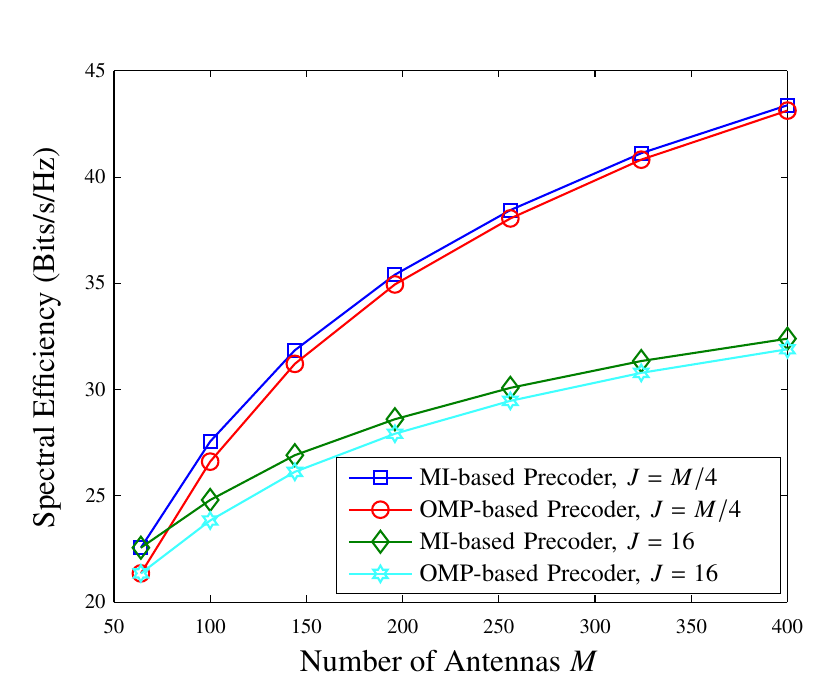}
	\caption{Spectral efficiency (bits/s/Hz) versus number of transmit antennas $M$ for SI, $Q=N=4$, and $J\in\{16,M/4\}$. \vspace{-0.3cm}}
	\label{Fig:RateMPrecoder}
\end{figure}

In Fig.~\ref{Fig:RateMPrecoder}, we show the spectral efficiency $R$ (bits/s/Hz) from \eqref{Eq:RateMax} versus the number of transmit antennas $M$ for SI, $Q=N=4$,  and $J\in\{16,M/4\}$. Note that both IRS/ITS-aided antennas yield the same spectral efficiency since we neglect the impact of the blockage of the active antennas in  IRS-aided antennas and the difference in the array efficiency factor $\rho_A$ for these architectures only influences their power consumption but does not impact their spectral efficiency.  We observe that, as the number of antennas $M$ increases, the spectral efficiency increases. However, the slope of the increase is larger for $J=M/4$ than for $J=16$. As can be seen from Fig.~\ref{Fig:RateMPrecoder}, the MI-based procoder outperforms the OMP-based precoder in terms of spectral efficiency. This is expected since the MI-based procoder is optimized for maximization of the achievable rate whereas the OMP-based precoder is obtained by approximating the optimal unconstrained FD precoder. Nevertheless, the additional gain of the MI-based procoder is small and decreases as $M$ increases. This can be attributed to the fact that both the MI- and OMP-based precoders search over the same sets for their respective phase-shift matrices whose cardinality is rather small, i.e., $NL$. In fact, we show later in Fig.~\ref{Fig:RateL} that for rich scattering environments (i.e., large $L$), the proposed MI-based precoder achieves larger performance gains over the OMP-based precoder. For the remainder of this section (except for Fig.~\ref{Fig:RateL}), we consider poor scattering mmWave channels (specifically $L=8$), and hence focus on the OMP-based precoder.


\begin{figure}\vspace{-0.5cm}
	\centering
	\includegraphics[width=1\linewidth]{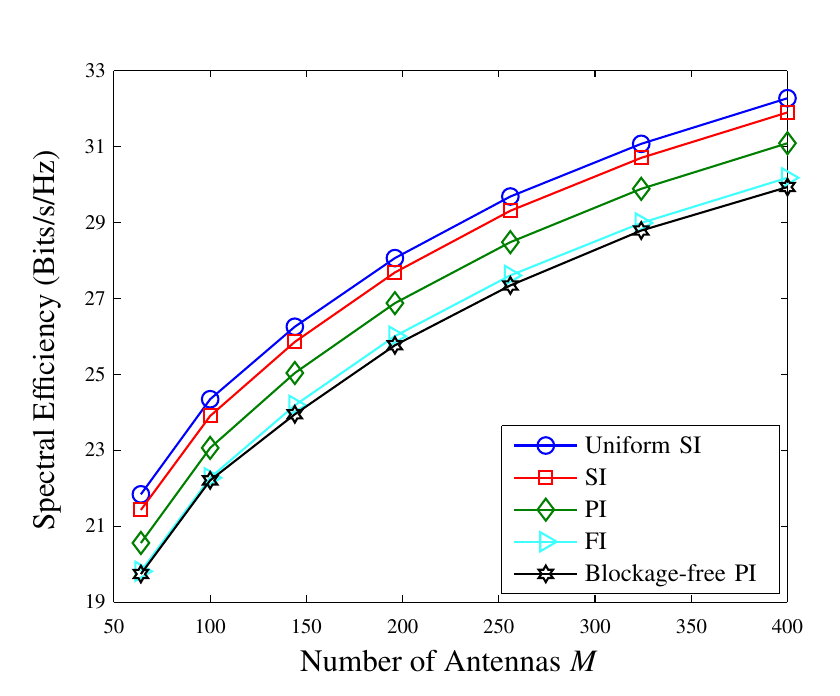}\vspace{-0.03cm}	
	\caption{Spectral efficiency (bits/s/Hz) versus number of transmit antennas $M$ for $Q=N=4$, $J=16$, and different illumination strategies. \vspace{-0.3cm}}
	\label{Fig:RateMIlluminarion}
\end{figure}

In Fig.~\ref{Fig:RateMIlluminarion}, we show the spectral efficiency (bits/s/Hz) versus the number of transmit antennas $M$ for $Q=N=4$ and $J=16$, and study the impact of the different illumination strategies introduced in Section~\ref{Sec:Illumination}, namely FI, PI, SI, blockage-free PI, and uniform SI, see Fig.~\ref{Fig:Illumination}. As can be observed from this figure, PI achieves a better performance than FI. This can be explained as follows. For the proposed precoder, the passive elements are partitioned into $N$ mutually exclusive subsets, i.e., $\mathcal{M}_n,\,\,n=1,\dots,N$, where each subset is responsible for reflection/transmission of the signal of one of the active antennas, cf. \eqref{Eq:Freform}. For PI, the positioning of the active antennas minimizes the interference between the different subsets of passive antennas $\mathcal{M}_n,\,\,n=1,\dots,N$, which is beneficial for performance, whereas for FI there is significant interference between different subsets of passive antennas, which cannot be mitigated by the precoder. To study the impact of the power distribution across the passive antennas, we also show the achievable rate for uniform SI, cf. Proposition~\ref{Prop:Precoder} and Corollary~\ref{Corol:Ideal}. As expected, uniform SI outperforms SI; nevertheless, the additional gain is small. Furthermore, we observe from Fig.~\ref{Fig:RateMIlluminarion} that blockage-free PI, which is designed to avoid the blockage of desired AoAs/AoDs by the active antennas for the IRS-aided architecture, achieves the lowest  spectral efficiency. This is due to the higher over-the-air power loss and the more non-uniform power distribution across the intelligent surface compared to the other illumination strategies.

\begin{figure} 
	\begin{minipage}{1\linewidth} 
		\centering
		\includegraphics[width=1.05\linewidth]{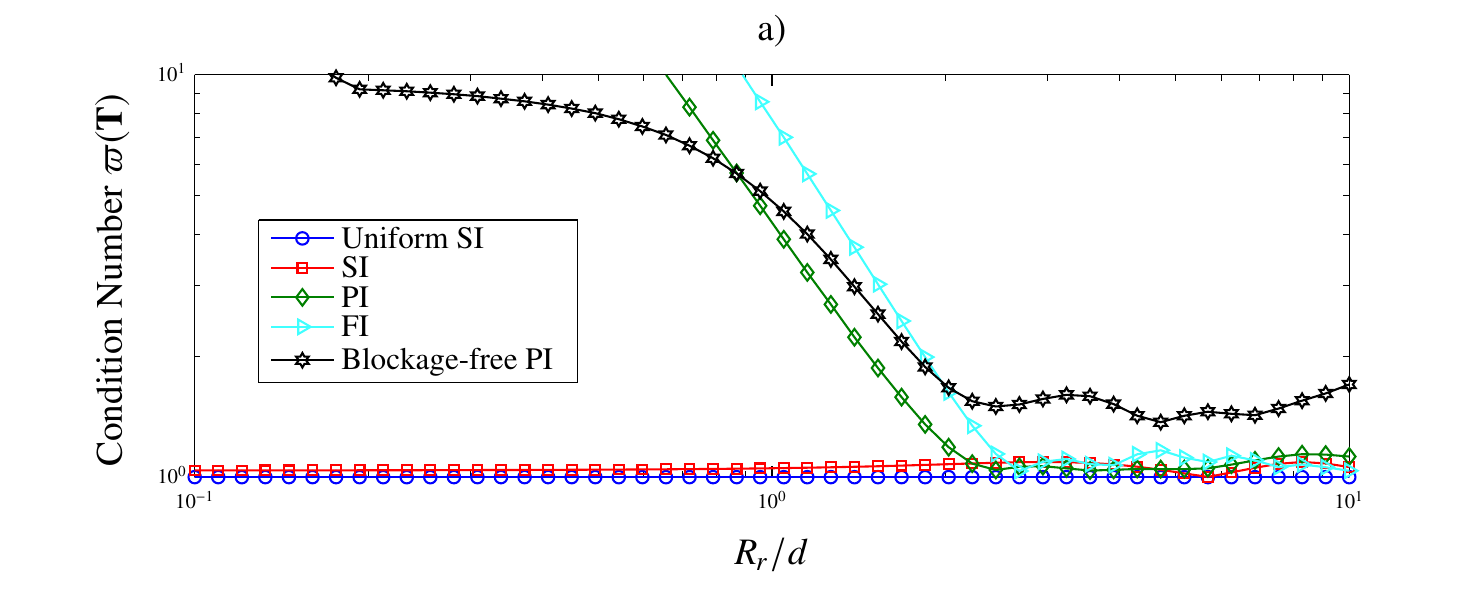}
	\end{minipage}	
	\begin{minipage}{1\linewidth} 
		\centering
		\includegraphics[width=1.05\linewidth]{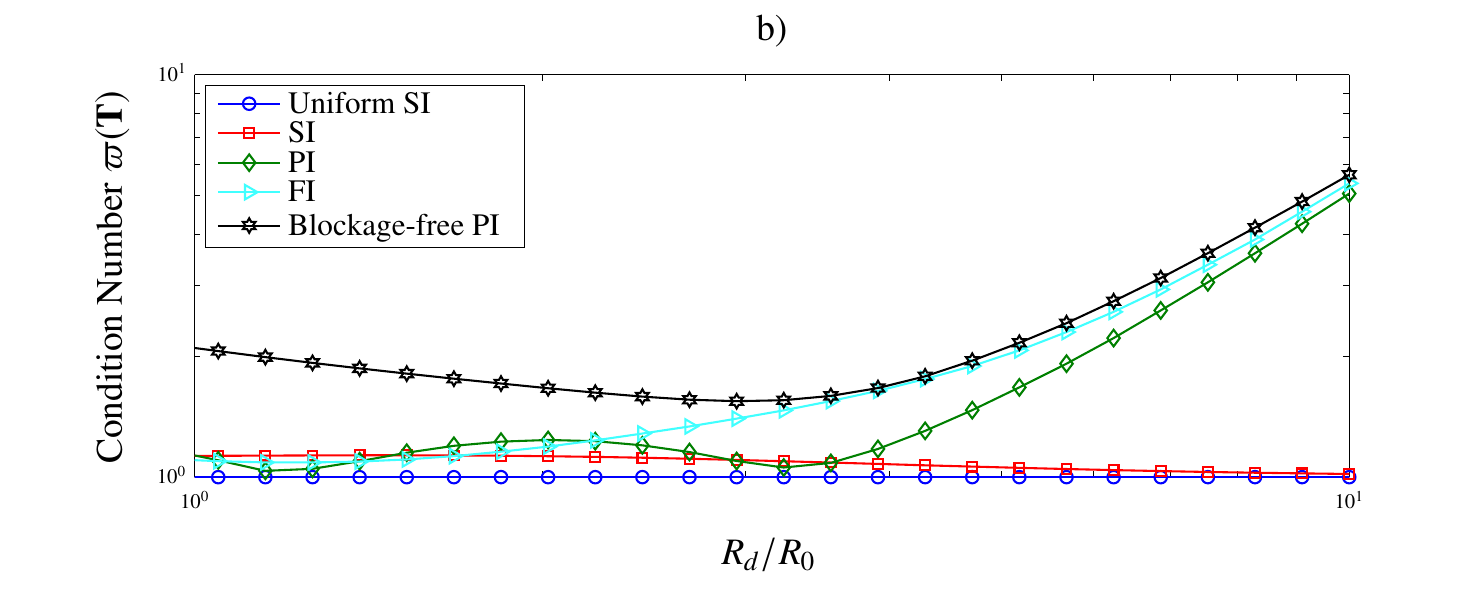}
	\end{minipage}		
	\caption{Condition number $\varpi(\mathbf{T})$ versus a) $R_r$ for $R_d=4R_0$ and b) $R_d$ for $R_r=2d$ assuming $M=256$, $N=4$, and different illuminations. \vspace{-0.03cm}}
	\label{Fig:Mmat}
\end{figure}

Next, we study the impact of the positioning of the active antennas and the intelligent surface via parameters $R_r$ and $R_d$. Unfavorable positioning of the active  antennas and the intelligent surface causes matrix $\mathbf{T}$ to be ill-conditioned which in turn decreases the achievable rate of any precoder design due to the reduced degrees of freedom in $\mathbf{F}=\mathbf{DTB}$. Moreover, an ill-conditioned matrix $\mathbf{T}$ leads to an increase of the power that has to be radiated by the active antennas to achieve a certain transmit power for the intelligent surface, {\BLUE see Lemma~\ref{Lem:PrdPtx}. Therefore, we study the condition number of matrix $\mathbf{T}$, i.e., $\varpi(\mathbf{T})$, for different illumination scenarios.} Recall that a  well-conditioned matrix $\mathbf{T}$ has a condition number close to one. In Fig.~\ref{Fig:Mmat}, we show the condition number $\varpi(\mathbf{T})$ versus a) $R_r$ for $R_d=4R_0$ where $R_0\triangleq d\sqrt{\frac{M}{\pi N}} $ (cf. (\ref{Eq:Rd})) and b) $R_d$ for $R_r=2d$ assuming $M=256$, $N=4$, and different illumination strategies, namely FI, PI, SI, blockage-free PI, and uniform SI. As can be seen from Fig.~\ref{Fig:Mmat} a), for FI, PI, and blockage-free PI, the condition number of $\mathbf{T}$ improves (i.e., decreases) as $R_r$ increases. Whereas for SI (uniform SI), the condition number of $\mathbf{T}$ is close to one (exactly one) for the entire considered range of $R_r$. On the other hand, Fig.~\ref{Fig:Mmat} b) shows that, for FI, PI, and blockage-free PI, the condition number of $\mathbf{T}$ generally increases as $R_d$ increases which is expected since the columns of matrix $\mathbf{T}$ become more similar. Interestingly for SI (uniform SI), the condition number of $\mathbf{T}$ remains again close to one (exactly one) for the entire considered range of $R_d$. From Figs.~\ref{Fig:RateMIlluminarion} and \ref{Fig:Mmat}, we conclude that SI yields  a better performance than FI and PI. This makes SI a suitable illumination option especially for the ITS-aided architecture which does not face the issue of the blockage of AoAs/AoDs by the active antennas. More importantly, as far as hardware implementation is concerned, SI is simpler than FI and PI since each active antenna and its respective passive elements can be  manufactured independent of the other active and passive antennas.

\subsection{Comparison of Different MIMO Architectures}

\begin{figure}
		\centering\vspace{-0.05cm}
	\includegraphics[width=1\linewidth]{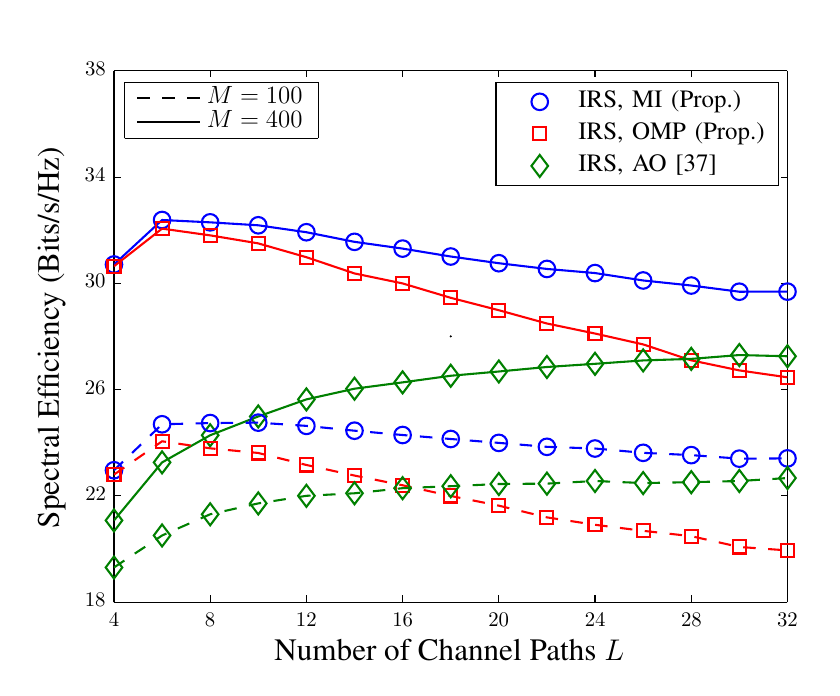}
		\caption{Spectral efficiency (bits/s/Hz) versus number of channel paths $L$ for $Q=N=4$, $M\in\{100,400\}$, and $J\in\{16,36\}$. \vspace{-0.03cm}}
		\label{Fig:RateL}
\end{figure}

For IRS/ITS-aided antennas, we adopt the proposed OMP-based precoder (except for Fig.~\ref{Fig:RateL}) and the SI illumination strategy. For comparison, in addition to the benchmark schemes discussed in Section~\ref{Sec:Benchmark}, we consider the AO-based precoder in \cite{zhou2018hardware} for IRS-aided antennas. In this case, we adopt the FI strategy, not SI, since this precoder was not designed for SI and, as a result, has a poor performance in this case.

\begin{figure}
		\centering\vspace{-0.05cm}
		\includegraphics[width=1\linewidth]{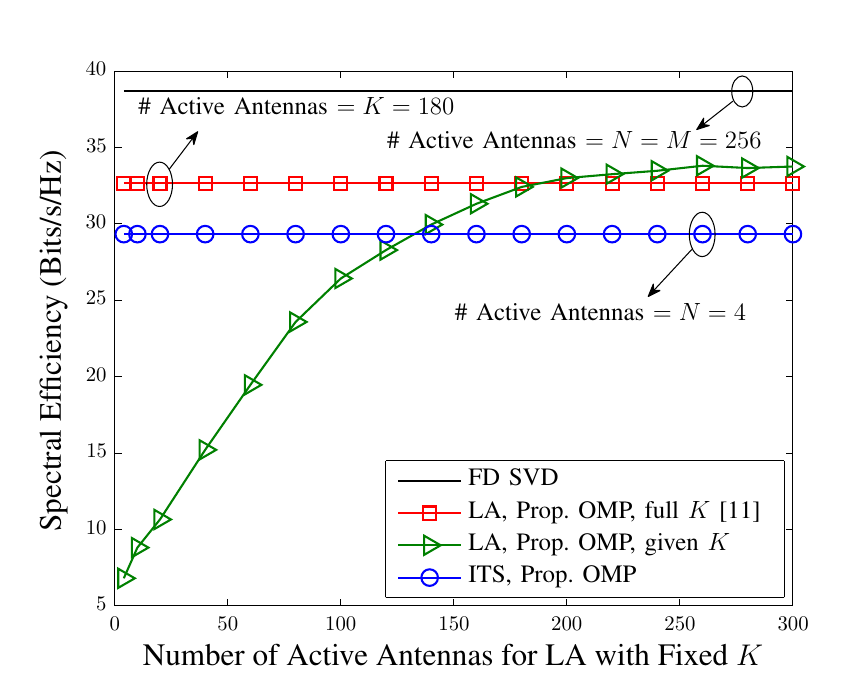}
		\caption{Spectral efficiency (bits/s/Hz) versus number of active antennas $K$ in the LA MIMO architecture for $Q=N=4$, $M=256$, and $J=16$. \vspace{-0.03cm}}
		\label{Fig:RateK}
\end{figure}

In Fig.~\ref{Fig:RatePowerEE_M}, we show a) the spectral efficiency $R$ (bits/s/Hz) given in (\ref{Eq:RateMax}), b) the corresponding total consumed power $\Ptot$ (dBm), and c) the corresponding energy efficiency $WR/\Ptot$ (bits/joule) versus the number of transmit antennas $M$ for $Q=N=4$ and $J=16$.  Our observations from Fig.~\ref{Fig:RatePowerEE_M} can be summarized as follows:

\begin{itemize}
\item As can be seen from Fig.~\ref{Fig:RatePowerEE_M}~a), the FC hybrid architecture can closely approach the spectral efficiency of the FD architecture. As expected, PC hybrid MIMO  has a lower spectral efficiency compared to FC hybrid MIMO  due to the fewer degrees of freedom available for beamforming in PC MIMO, i.e., only $M$ phase shifters are used in PC MIMO whereas $MN$ phase shifters are employed in FC MIMO. Although the IRS/ITS-aided  architectures also have $M$ phase shifters, they achieve a slightly lower spectral efficiency compared to the PC architecture due to non-uniform power distribution across the intelligent surface.	The LA antennas with full $K$ outperform the PC and IRS/ITS-aided  antennas since the entire surface/lens can be used for transmission of the signal from each active antenna due to their placement on the focal arc of the lens. However, in practice, linearly increasing $K$ with $M$ is infeasible as active antennas are costly and bulky. For a fixed $K$ of 64, the achievable rate of the LA antenna even decreases with $M$ which is due the fixed number of supported AoDs and the narrow beam generated by the lens. We investigate the impact of $K$ on the performance of LA antennas in more detail in Fig.~\ref{Fig:RateK}.
\item We also observe from Fig.~\ref{Fig:RatePowerEE_M} a) that  the proposed OMP-based precoder outperforms the AO-based precoder in \cite{zhou2018hardware} by a large margin. This might be attributed to the fact that the iterative AO-based algorithm in \cite{zhou2018hardware} is more prone to getting trapped in a local optimum which is avoided by the proposed OMP-based precoder as it efficiently exploits the sparsity of  mmWave channels. 



\begin{figure}[!t]
	\begin{minipage}{1.02\linewidth}
		\centering\vspace{-0.05cm}
		\includegraphics[width=1\linewidth]{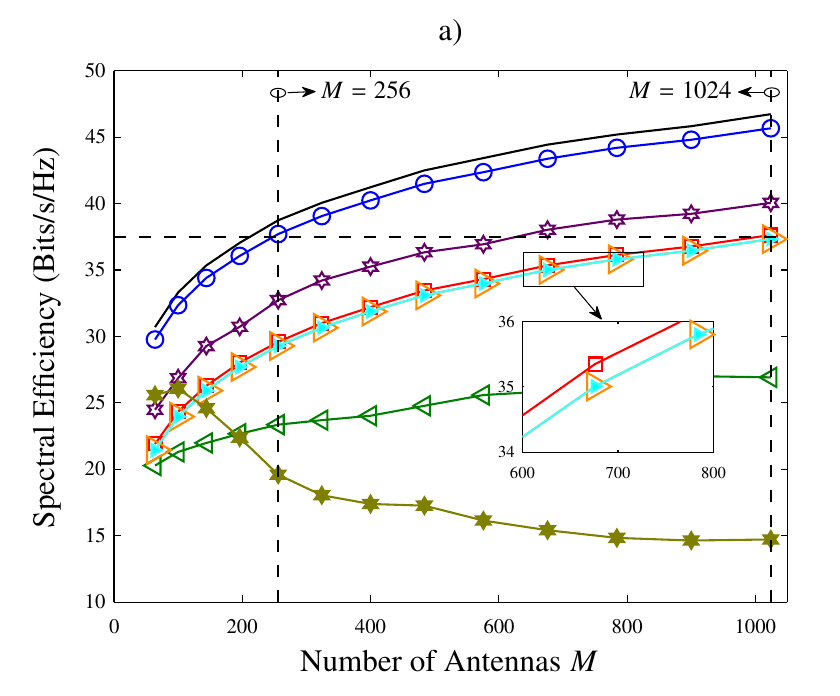}\vspace{-0.2cm}
	\end{minipage}
	\begin{minipage}{1.02\linewidth}
		\centering\vspace{-0.05cm}
		\includegraphics[width=1\linewidth]{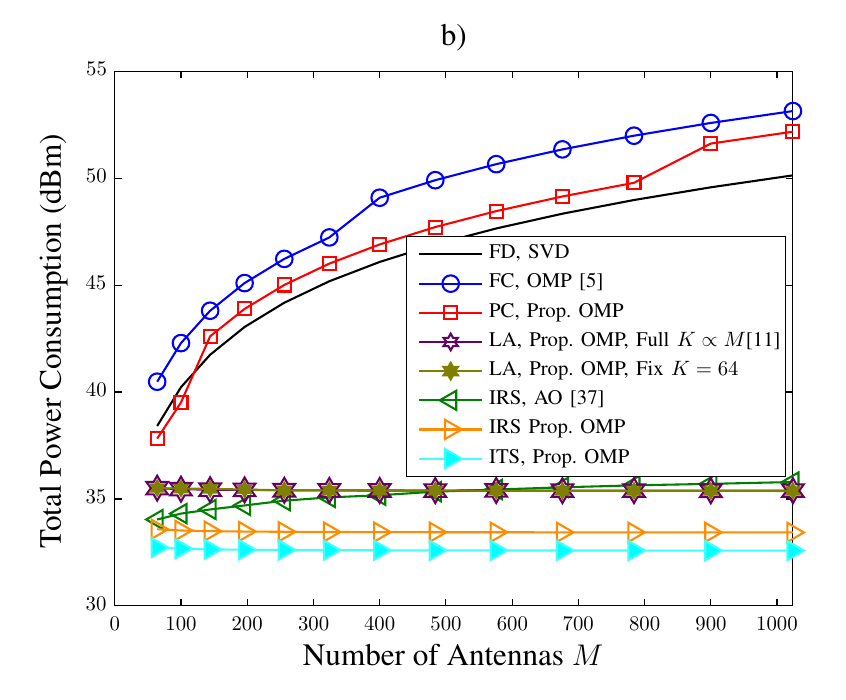}
	\end{minipage}
	\begin{minipage}{1.02\linewidth}
		\centering\vspace{-0.05cm}
		\includegraphics[width=1\linewidth]{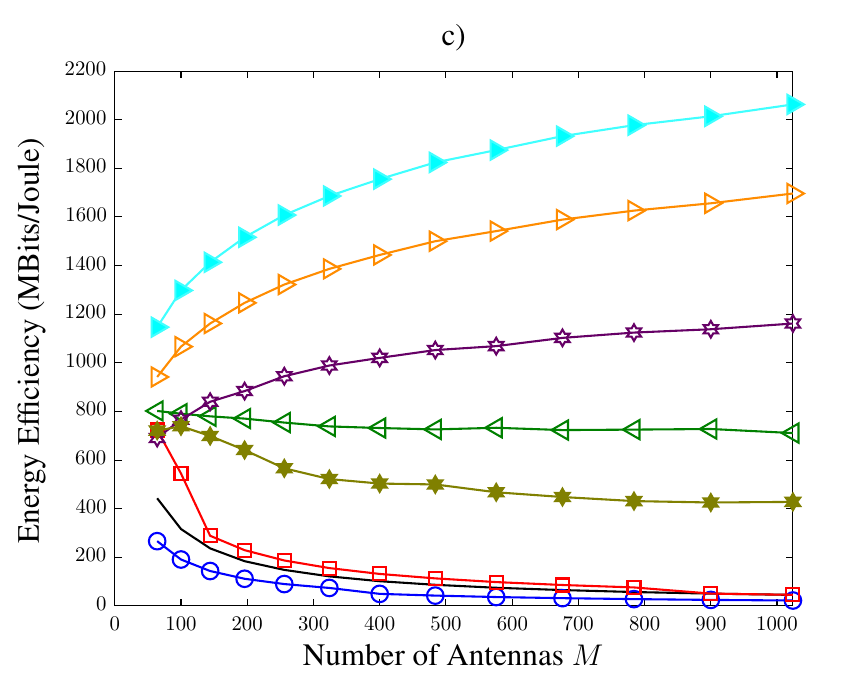}
	\end{minipage}
	\caption{a) Spectral efficiency (bits/s/Hz), b) total consumed power $\Ptot$ (dBm), and c) energy efficiency (MBits/Joule) versus number of transmit antennas $M$ for   $Q=N=4$ and $J=16$. }
	\label{Fig:RatePowerEE_M}
\end{figure}

\item The main advantage of the IRS/ITS-aided architectures is their scalability in terms of the number of antennas $M$ which is evident from Figs.~\ref{Fig:RatePowerEE_M} b) and c). In fact, IRS/ITS-aided MIMO (using the proposed precoder) achieve similar spectral efficiency as FD and FC MIMO if they are equipped with $N$ times more antennas, e.g., in Fig.~\ref{Fig:RatePowerEE_M} a), FD and FC MIMO with $M=256$ antennas and IRS/ITS-aided MIMO with $M=1024$ antennas achieve the same spectral efficiency of $37.5$ bits/s/Hz. However, from  Fig.~\ref{Fig:RatePowerEE_M} b), we observe that the total transmit power of the conventional FD, FC, and PC architectures significantly increases as $M$ increases which makes their implementation quite costly or even infeasible\footnote{We note that the jumps in the power consumptions of the FC and PC MIMO architectures in Fig.~\ref{Fig:RatePowerEE_M} b) are due to an increase of the number in required GCAs per antenna, i.e., $\lceil\frac{L_{\mathrm{rf}}}{G_{\mathrm{amp}}}\rceil$, as $M$ increases.}. {\BLUE We note that the high power consumption of FD antennas can be attributed to the large number of DACs whereas that of the PC and FC hybrid antennas is mainly caused by the GCAs needed to compensate for the power loss in the analog network  \cite{yan2018performance}.} On the other hand, the total power consumption of the LA and IRS/ITS-aided architectures remains almost constant as $M$ increases, {\BLUE which is mainly due to the efficient  over-the-air connection between the active and passive antennas \cite{abdelrahman2017analysis,nayeri2018reflectarray}.}

\item We observe from Fig.~\ref{Fig:RatePowerEE_M} b) that the power consumption of the LA architecture does not depend on the number of active antennas $K$. However, the overall power consumption of LA antennas is higher than that of IRS/ITS-aided antennas since the active antennas have to transmit with higher power $\Prd$ to achieve the same transmit power of the intelligent surface $\Ptx$. In particular, for the LA architecture, the active antennas have to be placed on the focal arc of the lens, which leads to a severely non-uniform power distribution across the intelligent surface for the received signal of some of the active antennas. This issue does not exist for IRS/ITS-aided antennas which is the reason for their lower power consumption compared to LA antennas. In addition, from Fig.~\ref{Fig:RatePowerEE_M} b), we observe that the proposed ITS-aided antennas have a lower power consumption compared to the proposed IRS-aided antennas due to their higher array efficiency factor, i.e., $[\rho_{\mathrm{srf}}]_{\text{dB}}=2[\rho_P]_{\text{dB}}+[\rho^{\rm srf}_A]_{\text{dB}}=-4.5$~dB and $[\rho_{\mathrm{srf}}]_{\text{dB}}=[\rho_P]_{\text{dB}}+[\rho^{\rm srf}_A]_{\text{dB}}=-3.5$~dB, cf. Table~I.  Furthermore, the precoder in \cite{zhou2018hardware} leads to a higher power consumption  than the proposed precoder since the latter employs SI whereas the former~employs~FI. 

\item We observe in Fig.~\ref{Fig:RatePowerEE_M} c) that the energy efficiency of the conventional  FD, FC, and PC architectures decreases as  $M$ increases whereas the energy efficiency of the proposed IRS/ITS-aided architectures increases.  This is mainly due to the high power consumption of the conventional architectures for large $M$. For LA antennas with fixed $K=64$ (full $K$), the energy efficiency decreases (increases) due to the decreasing (increasing) spectral efficiency as $M$ increases.
\end{itemize}

In Fig.~\ref{Fig:RateK}, we show the spectral efficiency (bits/s/Hz) of the LA architecture versus the number of active antennas $K$ for $Q=N=4$, $M=256$, and $J=16$. In addition, we also show the performance of the LA architecture with the full $K$ design in  \cite{zeng2018multi} ($K=180$ for $M=256$ and the considered range of AoDs), the proposed ITS-aided architecture (with $4$ active antennas), and the FD architecture (with $256$ active antennas). As expected the achievable rate of LA antennas improves with $K$ due to the larger number of supported AoDs. However, the curve saturates for large $K$ (approximately $K>M$) since in this regime, the bottleneck is the passive lens (i.e., how narrow the beam can be made). Fig.~\ref{Fig:RateK} shows that at $M=256$, the proposed ITS-aided architecture with only $4$ active antennas outperforms the LA architecture with $K=132$ antennas.

In Fig.~\ref{Fig:RateL}, we compare the performance of the AO-based precoder in \cite{zhou2018hardware} and the proposed MI-based and OMP-based precoders in more detail for different scattering environments and different numbers of transmit antennas. In particular, in Fig.~\ref{Fig:RateL}, we show the spectral efficiency (bits/s/Hz) versus the number of channel paths $L$ for $Q=N=4$, $M\in\{100,400\}$, and $J=16$. From this figure, we observe  that as the number of channel paths $L$ increases, the spectral efficiency of the proposed MI-based and OMP-based precoders first increases and then decreases.  This behavior is due to the fact that the proposed precoders select the $N$ best paths. Therefore, for larger $L$, we have more paths to select from, which yields a diversity gain, but selecting only $N$ out of $L$ paths becomes a limiting factor.  In contrast, the spectral efficiency of the AO-based precoder in \cite{zhou2018hardware} increases as $L$ increases. This is due to the fact that the AO-based precoder in \cite{zhou2018hardware} does not explicitly choose its phase-shift matrix based on the transmit array response of the available paths. In fact, increasing $L$ leads to a better conditioned channel matrix which improves the convergence behavior of the precoder  in \cite{zhou2018hardware}. In addition, we observe from Fig.~\ref{Fig:RateL} that the performance gain of the proposed precoder over the AO-based precoder increases with the number of transmit and receive antennas. This behavior is in line with the results reported in the literature which state that for large $M$, even the optimal unconstrained precoder transmits the data over at most the $N$ strongest channel paths  \cite{el2014spatially,ghanaatian2018feedback}. Finally, we observe that for all ranges of the parameters considered in Fig.~\ref{Fig:RateL}, the proposed MI-based precoder outperforms both the proposed OMP-based precoder and the AO-based precoder in \cite{zhou2018hardware}.

\subsection{CSI Imperfection}
\begin{figure}
	\centering
	\includegraphics[width=1\linewidth]{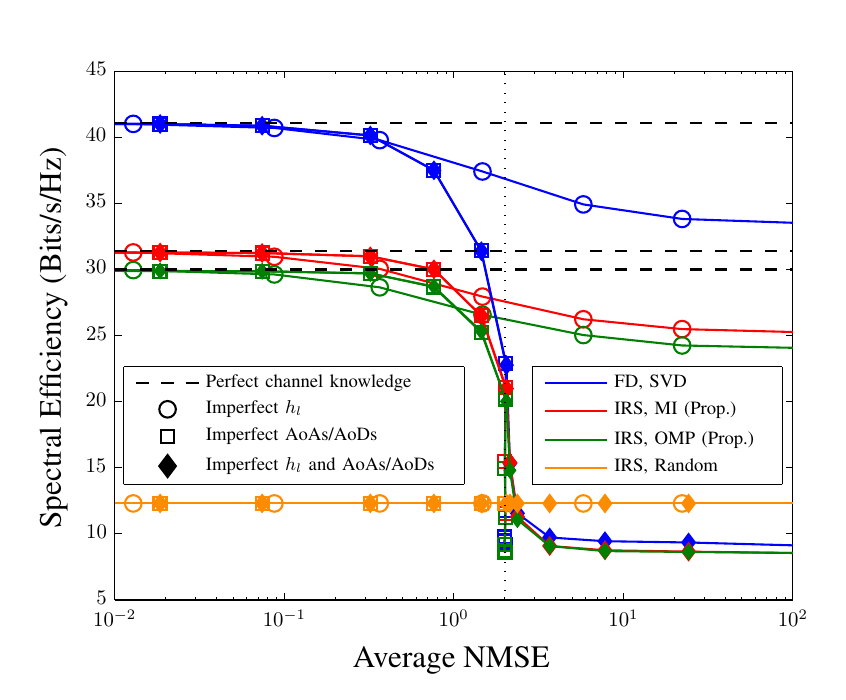}
	\caption{\BLUE Spectral efficiency (bits/s/Hz) versus average NMSE for $M=400$, $Q=N=4$, $J=16$, and $L=16$. }
	\label{Fig:RateCSI}
\end{figure}

Next, we investigate the impact of channel estimation errors on the performance of the proposed precoder design. We assume that the channel is reconstructed by estimating the AoAs $(\theta_l^r,\phi_l^r)$, the AoDs $(\theta_l^t,\phi_l^t)$, and the corresponding channel coefficients $h_l$. The estimated parameters are modeled as  $\hat{x} = x+\epsilon_x,\,\,x\in\{\theta_l^r,\phi_l^r,\theta_l^t,\phi_l^t,h_l\}$, where $\epsilon_x$ denotes the estimation error and is a (real for AoAs/AoDs and complex for $h_l$) zero-mean Gaussian RV.  For simplicity, we assume that the estimation errors of different parameters are independent, the estimation error variances for the AoAs and AoDs are identical, and the number of paths $L$ is correctly estimated. As a measure for the estimation quality, we use the widely-adopted normalized mean square error (NMSE), defined as $\|\hat{\mathbf{H}}-\mathbf{H}\|_F^2/\|\mathbf{H}\|_F^2$, where $\hat{\mathbf{H}}$ denotes the estimated channel matrix \cite{lee2016channel,zhao2017multi}. We use the estimated channel matrix $\hat{\mathbf{H}}$ for the precoder design, but compute the spectral efficiency for the actual channel matrix $\mathbf{H}$. Fig.~\ref{Fig:RateCSI} shows the spectral efficiency of the FD and IRS-aided architectures versus the NMSE for $M=400$, $Q=N=4$, and $J=16$. To separately investigate the impact of estimation errors of the AoAs/AoDs and the path coefficients, we consider {\BLUE three} cases, namely \textit{i)} the AoAs and AoDs are estimated and the path coefficients $h_l$ are known,  \textit{ii)} the path coefficients $h_l$ are estimated and the AoAs and AoDs are known, and {\BLUE \textit{iii)} the path coefficients $h_l$ and  the AoAs/AoDs are estimated and the estimation error $\epsilon_x$ has the same variance, $\forall x\in\{\theta_l^r,\phi_l^r,\theta_l^t,\phi_l^t,h_l\}$. In addition to FD antennas and IRS-aided antennas using MI-based and OMP-based precoders,  we also show results for IRS-aided antennas with a random precoder as a baseline. The random precoder $\F=\D\T\B$ is computed by initially generating the entries of $\D$ and $\B$ as zero-mean unit-variance complex Gaussian RVs, then normalizing each entry of $\D$ to ensure the unit-modulus constraint, and finally normalizing matrix $\B$ to ensure the transmit power constraint $\|\F\|_F=1$, such that the random precoder satisfies $\F\in\mathcal{F}$.} From Fig.~\ref{Fig:RateCSI}, we observe that for small NMSE (approx. below $0.1$), the performances of the considered precoders designed based on estimated channel $\hat{\mathbf{H}}$ and perfect channel $\mathbf{H}$ are similar. For the estimated AoAs/AoDs, the achievable spectral efficiency decreases as the NMSE increases. Note that, in this case, the NMSE is upper bounded by $2$. Here, the upper bound corresponds to the extreme case where the estimated and the actual AoAs/AoDs are independent RVs, which yields $\Ex\{\|\hat{\mathbf{H}}-\mathbf{H}\|_F^2\}=2\Ex\{\|\mathbf{H}\|_F^2\}$.   In contrast,  for  estimated $h_l$, although the achievable spectral efficiency decreases as NMSE increases, it saturates to a larger value. The reason for this behavior is that the values of $h_l$ determine the $N$ best directions among the known AoDs/AoAs for transmission and the corresponding power allocation to the beams. Since the number of paths is small, i.e., $L=16$, and the actual AoAs/AoDs are known,  we can achieve a large rate even by random path selection and random power allocation where the randomness is due to severely imperfect $\hat{h}_l$. {\BLUE On the other hand, if the AoDs are incorrectly estimated, the precoder transmits the signal in wrong directions which severely reduces the amount of signal power that reaches the receiver. In fact, for \BLUE large NMSEs,  precoder designs based on  imperfect CSI are even outperformed by the random precoder which scatters the signal in all directions. Moreover, we observe from Fig.~\ref{Fig:RateCSI}  that the imperfect AoAs/AoDs are the performance bottleneck when both the AoAs/AoDs and the path coefficients are imperfect.} Furthermore, Fig.~\ref{Fig:RateCSI} suggests that the impact of channel estimation errors on the achievable spectral efficiency of the FD and IRS-aided architectures is quite similar.

\section{Conclusions and Future Work}\label{Sec:Cncln}
In the following, we first briefly present the major conclusions drawn from this paper, and then we provide some directions for future research.

\subsection{Conclusions} 

In this paper, we proposed to employ IRS and ITS to realize the full potential of mmWave \textit{ultra massive} MIMO in practice. In particular, we derived models for the corresponding precoder structure and the consumed power that accounted for the imperfections of IRS/ITS. Furthermore, we proposed different illumination strategies for the active antennas to realize the full potential of IRS/ITS-aided MIMO. Based on the derived precoder structure and exploiting the sparsity of mmWave channels, we designed two efficient precoders for IRS/ITS-aided antennas; namely MI- and OMP-based precoders. Our  comprehensive simulation studies provided the following  interesting insights for system design: 1) A proper positioning of the active antennas with respect to the intelligent surface leads to a considerable improvement in spectral efficiency of IRS/ITS-aided MIMO architectures. 2) {\BLUE In poor scattering environments, the proposed OMP-based precoder approaches the performance of the proposed MI-based precoder, whereas in rich scattering environments, there is a performance gap between these two precoders. Therefore, in channels with poor scattering, the proposed OMP-based precoder is preferable due to its lower computational complexity, whereas in channels with rich scattering, the proposed MI-based precoder may be preferable due to its superior performance.}   3) For large numbers of transmit antennas, the conventional MIMO architectures are either energy inefficient (due to the high power consumption of the large numbers of RF chains of the FD MIMO architecture and the large power loss in the RF network of the hybrid MIMO architectures) or expensive and bulky (due to the large numbers of active antennas required for the LA MIMO architecture). 4) The proposed IRS/ITS-aided MIMO architectures are highly energy efficient (because of the almost lossless over-the-air connection between the active antennas and the passive intelligent surface) and fully scalable in terms of the number of transmit antennas (since unlike for LA antennas, only few active antennas are required). 

\subsection{Future Work}

The results of this paper can be extended in several directions:
	
	\textbf{Reception Design:}  
	In modern systems, the same node usually serves as both transmitter and receiver, e.g., a base station (BS) in cellular networks. Since the resources used for transmission and reception are typically orthogonal in time and/or frequency, it is customary to treat the transmission and reception problems separately. In this paper, we studied the use of IRS/ITS-aided antennas at the transmitter and focused on the modeling and design of the precoder. An important direction for future research  is to investigate the application of IRS/ITS-aided antennas at the receiver and to develop corresponding combining schemes.

\textbf{Multi-user Communications:} In this paper, we considered a point-to-point MIMO
	system, which is a suitable model for e.g. data backhauling
	from a macro BS to a small-cell BS. Other
	interesting network architectures include the uplink
	and downlink between a BS and multiple mobile users. The
	design of corresponding precoders and combiners when the BS
	employs IRS/ITS-aided antennas constitutes an interesting research
	problem for future work, see \cite{lin2016energy}. Hereby, one may consider different precoder design criteria including maximization of the users' sum rate, fairness among the users, and the coverage~area.

\textbf{Channel Estimation:} In this paper, we assumed that knowledge of channel matrix $\mathbf{H}$ is available at both transmitter and receiver, which is a widely-adopted assumption for precoder design in the literature \cite{el2014spatially,gao2016energy,yu2016alternating,zhou2018hardware,alkhateeb2016frequency,ribeiro2018energy,mirza2018joint,zang2019optimal,zeng2019energy,lin2016energy,castanheira2017hybrid}. In practice, channel matrix $\mathbf{H}$ has to be estimated from pilot symbols. A simple approach is based on beam steering whereby the transmitter and the receiver create pencil beams along specific AoDs $(\theta_l^t,\phi_l^t)$ and AoAs $(\theta_l^r,\phi_l^r)$, respectively, and estimate the corresponding channel coefficient $h_l$. Channel matrix $\mathbf{H}$ can be estimated by repeating this procedure for all possible AoDs and AoAs with a predefined resolution. Since this simple approach may cause a huge training overhead for high-quality channel estimation, more efficient channel estimation techniques have been developed in \cite{alkhateeb2014channel,zhao2017multi,delgado2018feasible} using compressed sensing, which exploits the sparsity of the mmWave channel. Although the beam steering technique can be directly applied to IRS/ITS-aided antennas, the existing compressed sensing-based schemes rely on the precoder structure of the conventional hybrid beamforming architectures, and hence, have to be suitably modified for application to IRS/ITS-aided antennas. Moreover, since the number of pilot symbols is limited due to the finite coherence time of the channel, estimation errors are unavoidable. This motivates the design of robust precoders for IRS/ITS-aided architectures which account for estimation errors, see \cite{roth2018comparison} for robust precoder designs for conventional hybrid and FD architectures.

	\appendices

{\BLUE
\section{Proof of Lemma~\ref{Lem:PrdPtx}}\label{App:LemPrdPtx}

The power radiated by the active antennas is obtained as $\Prd=\Tr\big\{\Ex\{\bar{\mathbf{x}}\bar{\mathbf{x}}\Herm\}\big\}\overset{(a)}{=}\Ptx\Tr\{\B\B\Herm\}=\Ptx\|\B\|_F^2$, where equality $(a)$ follows from $\bar{\mathbf{x}}=\sqrt{\Ptx}\B\mathbf{s}$ and $\Ex\{\mathbf{s}\mathbf{s}\Herm\}=\mathbf{I}_Q$. Similarly, the transmit power radiated by the intelligent surface is given by $\Ptx=\Tr\big\{\Ex\{\mathbf{x}\mathbf{x}\Herm\}\big\}\overset{(a)}{=}\Ptx\Tr\{\F\F\Herm\}\overset{(b)}{=}\Ptx\Tr\{\D\T\B\B\Herm\T\Herm\D\Herm\}\overset{(c)}{=}\Ptx\Tr\{\T\B\B\Herm\T\Herm\D\Herm\D\}=\Ptx\|\T\B\|_F^2$, where equality $(a)$ follows from $\bar{\mathbf{x}}=\sqrt{\Ptx}\F\mathbf{s}$ in \eqref{Eq:Sig_Gen}, equality $(b)$ follows from $\F=\D\T\D$ from Proposition~\ref{Prop:Precoder}, and equality $(c)$ follows from the matrix equality $\Tr\{\mathbf{A}\mathbf{B}\}=\Tr\{\mathbf{B}\mathbf{A}\}$. Therefore, $\|\T\B\|_F^2=1$ has to hold. Moreover, the Frobenius norm of the product of two matrices are bounded as \cite{bodewig2014matrix}
\begin{IEEEeqnarray}{lll}\label{Eq:Frbound}
	\sigma^2_{\min}(\T)\|\B\|_F^2\leq\|\T\B\|_F^2\leq\sigma^2_{\max}(\T)\|\B\|_F^2.
\end{IEEEeqnarray}
Using the above bounds and recalling that $\|\T\B\|_F^2=1$ holds, we obtain  bounds $\sigma^{-2}_{\max}(\T)\leq\|\B\|_F^2\leq\sigma^{-2}_{\min}(\T)$ whose substitution in $\Prd=\Ptx\|\B\|_F^2$ leads to the tighter lower and upper bounds in \eqref{Eq:PowerPtxPrd}.
Moreover, $\sigma^2_{\min}(\T)$ and $\sigma^2_{\max}(\T)$ are bounded respectively as
\begin{IEEEeqnarray}{lll} \label{Eq:Tminmax}
	\|\T\|_F^2&=\sum_n\sigma_n^2(\T) \nonumber\\
	&= \sigma^2_{\min}(\T) \sum_n \frac{\sigma_n^2(\T)}{\sigma^2_{\min}(\T)} \leq N \sigma^2_{\min}(\T)\varpi^2(\mathbf{T})\quad \IEEEyesnumber\IEEEyessubnumber\\
	\|\T\|_F^2&=\sum_n\sigma_n^2(\T) \nonumber\\
	&= \sigma^2_{\max}(\T) \sum_n \frac{\sigma_n^2(\T)}{\sigma^2_{\max}(\T)} \geq \frac{N\sigma^2_{\max}(\T)}{\varpi^2(\mathbf{T})}.\IEEEyessubnumber
\end{IEEEeqnarray}
On the other hand, $\|\T\|_F^2$ can be computed as
\begin{IEEEeqnarray}{lll}\label{Eq:Tnorm}
	\|\T\|_F^2&=\sum_n\rho_{\rm srf} \nonumber\\
	&\,\,\,\,\times\underset{A_n}{\underbrace{\sum_m\left(\frac{\lambda}{4\pi r_{m,n}}\right)^2 G^a(\theta^p_{m,n},\phi^p_{m,n}) G^p(\theta^a_{m,n},\phi^a_{m,n})}} \nonumber \\
	&=N\rho_{\rm srf}\rho_A^{\rm ant}\rho_S,\quad
\end{IEEEeqnarray}
where $A_n$ is the fraction of power radiated from active antenna $n$ and collected by the surface. $A_n$ is related to spillover loss $\rho_S$ and the passive antenna aperture loss $\rho_A^{\rm ant}$. Assuming that $\rho_A^{\rm ant}$ and $\rho_S$ are identical for all active antennas, we obtain $A_n=\rho_A^{\rm ant}\rho_S,\,\,\forall n$. Substituting \eqref{Eq:Tnorm} in \eqref{Eq:Tminmax} and then  in \eqref{Eq:Frbound} leads to the looser lower and upper bounds in \eqref{Eq:PowerPtxPrd} and concludes the proof.
}


	\section{Proof of Lemma~\ref{Lem:ArraySpace}}\label{App:Lem_ArraySpace}
	
	The proof follows from similar arguments as those provided in \cite{el2014spatially,molisch17,ghanaatian2018feedback,alkhateeb2016frequency} for the precoder design of conventional MIMO systems. In particular, for the achievable rate $R(\mathbf{F})$, the precoder appears in the term $\mathbf{HF}=\mathbf{H}\mathbf{F}_{\mathcal{H}_t}+\mathbf{H}\mathbf{F}_{\mathcal{H}_t^{\bot}}$. The elements of matrix $\mathbf{H}\mathbf{F}_{\mathcal{H}_t^{\bot}}$ are obtained based on the multiplication of the rows of $\mathbf{H}$ and the columns of $\mathbf{F}_{\mathcal{H}_t^{\bot}}$. Moreover, the rows of $\mathbf{H}$ belong to space $\mathcal{H}_t$ whereas the columns of $\mathbf{F}_{\mathcal{H}_t^{\bot}}$ belong to space $\mathcal{H}_t^{\bot}$. Since $\mathcal{H}_t$ and $\mathcal{H}_t^{\bot}$ are orthogonal, we obtain $\mathbf{HF}_{\mathcal{H}_t^{\bot}}=\boldsymbol{0}_{J,N}$. Therefore,  $\mathbf{F}_{\mathcal{H}_t^{\bot}}$ does not impact the achievable rate in (\ref{Eq:RateMax}). Moreover, we have $\|\mathbf{F}\|_F^2=\Tr(\mathbf{F}\mathbf{F}\Herm)=\Tr((\mathbf{F}_{\mathcal{H}_t}+\mathbf{F}_{\mathcal{H}_t^{\bot}})(\mathbf{F}_{\mathcal{H}_t}+\mathbf{F}_{\mathcal{H}_t^{\bot}})\Herm)=\Tr(\mathbf{F}_{\mathcal{H}_t}\mathbf{F}_{\mathcal{H}_t}\Herm+\mathbf{F}_{\mathcal{H}_t^{\bot}}\mathbf{F}_{\mathcal{H}_t^{\bot}}\Herm)=\|\mathbf{F}_{\mathcal{H}_t}\|_F^2+\|\mathbf{F}_{\mathcal{H}_t^{\bot}}\|_F^2\geq\|\mathbf{F}_{\mathcal{H}_t}\|_F^2$ where we used $\Tr(\mathbf{F}_{\mathcal{H}_t}\mathbf{F}_{\mathcal{H}_t^{\bot}}\Herm)=\Tr(\mathbf{F}_{\mathcal{H}_t^{\bot}}\mathbf{F}_{\mathcal{H}_t}\Herm)=0$. This concludes the proof.
	
	\section{Proof of Lemma~\ref{Lem:OptB}}\label{App:LemOptB}
	
	Let us define $\widetilde{\mathbf{B}}=\big(\mathbf{C}_1\Herm\mathbf{C}_1\big)^{\frac{1}{2}}\mathbf{B}\in\mathbb{C}^{Q\times N}$. Assuming $\mathbf{C}_1\Herm\mathbf{C}_1$ is a non-singular matrix, the constraint in \eqref{Eq:DigitalPrecoderRate} is rewritten  as
		\begin{IEEEeqnarray}{lll} 
			\Tr\left(\mathbf{C}_1\mathbf{B}\mathbf{B}\Herm \mathbf{C}_1\Herm\right) 
			\nonumber\\
			= \Tr\left(\mathbf{C}_1\big(\mathbf{C}_1\Herm\mathbf{C}_1\big)^{-\frac{1}{2}}\widetilde{\mathbf{B}}\widetilde{\mathbf{B}}\Herm\big(\mathbf{C}_1\Herm\mathbf{C}_1\big)^{-\frac{\mathsf{H}}{2}}\mathbf{C}_1\Herm\right) \nonumber\\
			\overset{(a)}{=} \Tr\left(\widetilde{\mathbf{B}}\widetilde{\mathbf{B}}\Herm\big(\mathbf{C}_1\Herm\mathbf{C}_1\big)^{-\frac{\mathsf{H}}{2}}\mathbf{C}_1\Herm\mathbf{C}_1\big(\mathbf{C}_1\Herm\mathbf{C}_1\big)^{-\frac{1}{2}}\right) \nonumber\\
			=\Tr\big(\widetilde{\mathbf{B}}\widetilde{\mathbf{B}}\Herm\big),
		\end{IEEEeqnarray}
	where for equality $(a)$, we used the relation $\Tr\left(\mathbf{X} \mathbf{Y}\right)=\Tr\left(\mathbf{Y} \mathbf{X}\right)$ for $\mathbf{X}\in\mathbb{C}^{n\times m}$ and $\mathbf{Y}\in\mathbb{C}^{m\times n}$. Based on this result and using the definition $\widetilde{\mathbf{H}}=\mathbf{H}\mathbf{C}_1\big(\mathbf{C}_1\Herm\mathbf{C}_1\big)^{-\frac{1}{2}}\in\mathbb{C}^{J\times N}$, the problem in \eqref{Eq:DigitalPrecoderRate} is rewritten as
		\begin{IEEEeqnarray}{cll}
			\underset{\widetilde{\mathbf{B}}\in\mathbb{C}^{N\times Q}}{\mathrm{maximize}} \,\,&\left|\mathbf{I}_{J}+\gamma \widetilde{\mathbf{H}}\widetilde{\mathbf{B}}\widetilde{\mathbf{B}}\Herm\widetilde{\mathbf{H}}\Herm\right| 
			\nonumber\\
			 \mathrm{subject\,\,to:}\,\, &\Tr\big(\widetilde{\mathbf{B}}\widetilde{\mathbf{B}}\Herm\big)\leq 1.
		\end{IEEEeqnarray}
	The above problem has the form of MI maximization for an FD MIMO system with equivalent channel matrix $\widetilde{\mathbf{H}}$. Thus, the solution is found via the waterfilling algorithm as $\widetilde{\mathbf{B}} = [\mathbf{v}_1,\dots,\mathbf{v}_Q]\mathbf{Z}$ \cite{ghanaatian2018feedback}. Then, the optimal baseband precoder is given by $\mathbf{B}=\big(\mathbf{C}_1\Herm\mathbf{C}_1\big)^{-\frac{1}{2}}\widetilde{\mathbf{B}}$ which concludes the proof.

\bibliographystyle{IEEEtran}
\bibliography{References}

\begin{thebibliography}{10}
\providecommand{\url}[1]{#1}
\csname url@samestyle\endcsname
\providecommand{\newblock}{\relax}
\providecommand{\bibinfo}[2]{#2}
\providecommand{\BIBentrySTDinterwordspacing}{\spaceskip=0pt\relax}
\providecommand{\BIBentryALTinterwordstretchfactor}{4}
\providecommand{\BIBentryALTinterwordspacing}{\spaceskip=\fontdimen2\font plus
\BIBentryALTinterwordstretchfactor\fontdimen3\font minus
  \fontdimen4\font\relax}
\providecommand{\BIBforeignlanguage}[2]{{%
\expandafter\ifx\csname l@#1\endcsname\relax
\typeout{** WARNING: IEEEtran.bst: No hyphenation pattern has been}%
\typeout{** loaded for the language `#1'. Using the pattern for}%
\typeout{** the default language instead.}%
\else
\language=\csname l@#1\endcsname
\fi
#2}}
\providecommand{\BIBdecl}{\relax}
\BIBdecl

\bibitem{jamali2018scalable}
V.~Jamali, A.~M. Tulino, G.~Fischer, R.~M{\"u}ller, and R.~Schober, ``{Scalable
  and Energy-Efficient Millimeter Massive MIMO Architectures: Reflect-Array and
  Transmit-Array Antennas},'' in \emph{Proc. IEEE ICC}, May 2019.

\bibitem{busari2017millimeter}
S.~A. Busari, K.~M.~S. Huq, S.~Mumtaz, L.~Dai, and J.~Rodriguez,
  ``{Millimeter-Wave Massive MIMO Communication for Future Wireless Systems: A
  Survey},'' \emph{IEEE Commun. Surveys \& Tutorials}, vol.~20, no.~2, pp.
  836--869, 2017.

\bibitem{delgado2018feasible}
A.~V. Delgado, M.~Sanchez-Fernandez, J.~Llorca, and A.~Tulino, ``{Feasible
  Transmission Strategies for Downlink MIMO in Sparse Millimeter-Wave
  Channels},'' \emph{IEEE Commun. Mag.}, vol.~56, no.~7, pp. 49--55, Jul. 2018.

\bibitem{gao2018low}
X.~Gao, L.~Dai, and A.~M. Sayeed, ``{Low RF-Complexity Technologies to Enable
  Millimeter-Wave MIMO with Large Antenna Array for 5G Wireless
  Communications},'' \emph{IEEE Commun. Mag.}, vol.~56, no.~4, pp. 211--217,
  Apr. 2018.

\bibitem{el2014spatially}
O.~El~Ayach, S.~Rajagopal, S.~Abu-Surra, Z.~Pi, and R.~W. Heath, ``{Spatially
  Sparse Precoding in Millimeter Wave MIMO Systems},'' \emph{IEEE Trans.
  Wireless Commun.}, vol.~13, no.~3, pp. 1499--1513, Mar. 2014.

\bibitem{molisch17}
A.~F. Molisch, V.~V. Ratnam, S.~Han, Z.~Li, S.~Le, H.~Nguyen, L.~Li, and
  K.~Haneda, ``{Hybrid Beamforming for Massive {MIMO}: A Survey},'' \emph{IEEE
  Commun. Mag.}, vol.~55, no.~9, pp. 134--141, Sep. 2017.

\bibitem{ghanaatian2018feedback}
R.~Ghanaatian, V.~Jamali, A.~Burg, and R.~Schober, ``{Feedback-Aware Precoding
  for Millimeter Wave Massive MIMO Systems},'' in \emph{Proc. IEEE PIMRC}, Sep.
  2019.

\bibitem{yan2018performance}
H.~Yan, S.~Ramesh, T.~Gallagher, C.~Ling, and D.~Cabric, ``{Performance, Power,
  and Area Design Trade-offs in Millimeter-Wave Transmitter Beamforming
  Architectures},'' \emph{IEEE Circuits Syst. Mag.}, vol.~19, no.~2, pp.
  33--58, second quarter 2019.

\bibitem{gao2016energy}
X.~Gao, L.~Dai, S.~Han, I.~Chih-Lin, and R.~W. Heath, ``{Energy-Efficient
  Hybrid Analog and Digital Precoding for mmWave MIMO Systems with Large
  Antenna Arrays},'' \emph{IEEE J. Select. Areas Commun.}, vol.~34, no.~4, pp.
  998--1009, Apr. 2016.

\bibitem{zeng2017cost}
Y.~Zeng and R.~Zhang, ``{Cost-Effective Millimeter-Wave Communications with
  Lens Antenna Array},'' \emph{IEEE Wireless Commun.}, vol.~24, no.~4, pp.
  81--87, Aug. 2017.

\bibitem{zeng2018multi}
Y.~Zeng, L.~Yang, and R.~Zhang, ``{Multi-User Millimeter Wave MIMO with
  Full-Dimensional Lens Antenna Array},'' \emph{IEEE Trans. Wireless Commun.},
  vol.~17, no.~4, pp. 2800--2814, Apr. 2018.

\bibitem{brady2013beamspace}
J.~Brady, N.~Behdad, and A.~M. Sayeed, ``{Beamspace MIMO for Millimeter-Wave
  Communications: System Architecture, Modeling, Analysis, and Measurements},''
  \emph{IEEE Trans. Antennas Propag.}, vol.~61, no.~7, pp. 3814--3827, Jul.
  2013.

\bibitem{popovic2002multibeam}
D.~Popovic and Z.~Popovic, ``{Multibeam Antennas with Polarization and Angle
  Diversity},'' \emph{IEEE Trans. Antennas Propag.}, vol.~50, no.~5, pp.
  651--657, May 2002.

\bibitem{balanis1982antenna}
C.~A. Balanis, ``{Antenna Theory, Analysis and Design},'' 1982.

\bibitem{wu2018intelligent}
Q.~Wu and R.~Zhang, ``{Intelligent Reflecting Surface Enhanced Wireless
  Network: Joint active and Passive Beamforming Design},'' in \emph{Proc. IEEE
  Globecom}, Dec. 2018, pp. 1--6.

\bibitem{liaskos2019novel}
C.~Liaskos, S.~Nie, A.~Tsioliaridou, A.~Pitsillides, S.~Ioannidis, and
  I.~Akyildiz, ``{A Novel Communication Paradigm for High Capacity and Security
  via Programmable Indoor Wireless Environments in Next Generation Wireless
  Systems},'' \emph{Ad Hoc Netw.}, vol.~87, pp. 1--16, 2019.

\bibitem{di2019smart}
{M. D. Renzo, et. al}, ``{Smart Radio Environments Empowered by AI
  Reconfigurable Meta-Surfaces: An Idea Whose Time Has Come},'' \emph{EURASIP
  J. Wireless Commun. Netw.}, vol. 129, May 2019.

\bibitem{basar2019wireless}
E.~Basar, M.~Di~Renzo, J.~De~Rosny, M.~Debbah, M.-S. Alouini, and R.~Zhang,
  ``Wireless communications through reconfigurable intelligent surfaces,''
  \emph{IEEE Access}, vol.~7, pp. 116\,753--116\,773, Sep. 2019.

\bibitem{wu2019intelligent}
Q.~Wu and R.~Zhang, ``{Intelligent Reflecting Surface Enhanced Wireless Network
  via Joint Active and Passive Beamforming},'' \emph{IEEE Trans. Wireless
  Commun.}, 2019.

\bibitem{najafi2019intelligent}
M.~Najafi and R.~Schober, ``{Intelligent Reflecting Surfaces for Free Space
  Optical Communications},'' Dec. 2019, pp. 1--7.

\bibitem{karasik2020beyond}
R.~Karasik, O.~Simeone, M.~Di~Renzo, and S.~Shamai, ``{Beyond Max-SNR: Joint
  Encoding for Reconfigurable Intelligent Surfaces},'' in \emph{ISIT}, May
  2020.

\bibitem{najafi2020intelligent}
\BIBentryALTinterwordspacing
M.~Najafi, V.~Jamali, R.~Schober, and H.~V. Poor, ``{Physical Modeling and
  Scalable Optimization of Large Intelligent Reflecting Surfaces},''
  \emph{under revision for IEEE Trans. Commun.}, 2020. [Online]. Available:
  \url{https://arxiv.org/abs/2004.12957}
\BIBentrySTDinterwordspacing

\bibitem{jamali2020power}
\BIBentryALTinterwordspacing
V.~Jamali, M.~Najafi, R.~Schober, and H.~V. Poor, ``{Power Efficiency,
  Overhead, and Complexity Tradeoff in IRS-Assisted Communications--Quadratic
  Phase-Shift Design},'' \emph{under revision for IEEE Commun. Lett.}, 2020.
  [Online]. Available: \url{https://arxiv.org/pdf/2009.05956}
\BIBentrySTDinterwordspacing

\bibitem{zuo2020intelligent}
J.~Zuo, Y.~Liu, E.~Basar, and O.~A. Dobre, ``{Intelligent Reflecting Surface
  Enhanced Millimeter-Wave NOMA Systems},'' \emph{arXiv preprint
  arXiv:2005.01562}, 2020.

\bibitem{abdelrahman2017analysis}
A.~H. Abdelrahman, F.~Yang, A.~Z. Elsherbeni, and P.~Nayeri, ``{Analysis and
  Design of Transmitarray Antennas},'' \emph{Synthesis Lectures on Antennas},
  vol.~6, no.~1, pp. 1--175, 2017.

\bibitem{luyen2019wideband}
H.~Luyen, Z.~Zhang, J.~H. Booske, and N.~Behdad, ``{Wideband, Beam-Steerable
  Reflectarrays Based on Minimum-Switch Topology, Polarization-Rotating Unit
  Cells},'' \emph{IEEE Access}, vol.~7, pp. 36\,568--36\,578, Apr. 2019.

\bibitem{popovic1998quasi}
Z.~Popovic and A.~Mortazawi, ``{Quasi-Optical Transmit/Receive Front Ends},''
  \emph{IEEE Trans. Microw. Theory Tech.}, vol.~46, no.~11, pp. 1964--1975,
  Nov. 1998.

\bibitem{di2015reconfigurable}
L.~Di~Palma, ``{Reconfigurable Transmitarray Antennas at Millimeter-Wave
  Frequencies},'' Ph.D. dissertation, University of Rennes, 2015.

\bibitem{pham2019low}
K.~T. Pham, A.~Clemente, E.~Fourn, F.~Diaby, L.~Dussopt, and R.~Sauleau,
  ``{Low-Cost Metal-Only Transmitarray Antennas at Ka-Band},'' \emph{IEEE
  Antennas Wireless Propag. Lett.}, vol.~18, no.~6, pp. 1243--1247, Jun. 2019.

\bibitem{hasani2019dual}
H.~Hasani, J.~S. Silva, S.~Capdevila, M.~Garcma-Vigueras, and J.~R. Mosig,
  ``{Dual-band Circularly Polarized Transmitarray Antenna for Satellite
  Communications at 20/30 GHz},'' \emph{IEEE Trans. Antennas Propag.}, vol.~67,
  no.~8, pp. 5325--5333, Aug. 2019.

\bibitem{bereyhi2019papr}
A.~Bereyhi, V.~Jamali, R.~R. M{\"u}ller, G.~Fischer, R.~Schober, and A.~M.
  Tulino, ``{PAPR-Limited Precoding in Massive MIMO Systems with Reflect-and
  Transmit-Array Antennas},'' in \emph{Asilomar Conf. Sig., Syst., and
  Computers}, Nov. 2019, pp. 1690--1694.

\bibitem{bereyhi2020single}
A.~Bereyhi, V.~Jamali, R.~R. M{\"u}ller, A.~M. Tulino, G.~Fischer, and
  R.~Schober, ``{A Single-RF Architecture for Multiuser Massive MIMO Via
  Reflecting Surfaces},'' in \emph{ICASSP}, May 2020, pp. 8688--8692.

\bibitem{hu2018beyond}
S.~Hu, F.~Rusek, and O.~Edfors, ``{Beyond Massive MIMO: The Potential of Data
  Transmission with Large Intelligent Surfaces},'' \emph{IEEE Trans. Sig.
  Process.}, vol.~66, no.~10, pp. 2746--2758, May 2018.

\bibitem{jung2020performance}
M.~Jung, W.~Saad, Y.~Jang, G.~Kong, and S.~Choi, ``{Performance Analysis of
  Large Intelligent Surfaces (LISs): Asymptotic Data Rate and Channel Hardening
  Effects},'' \emph{IEEE Trans. Wireless Commun.}, vol.~19, no.~3, pp.
  2052--2065, Mar. 2020.

\bibitem{nayeri2018reflectarray}
P.~Nayeri, F.~Yang, and A.~Z. Elsherbeni, \emph{{Reflectarray Antennas: Theory,
  Designs and Applications}}.\hskip 1em plus 0.5em minus 0.4em\relax Wiley
  Online Library, 2018.

\bibitem{wang2017spectrum}
B.~Wang, L.~Dai, Z.~Wang, N.~Ge, and S.~Zhou, ``{Spectrum and Energy-Efficient
  Beamspace MIMO-NOMA for Millimeter-Wave Communications Using Lens Antenna
  Array},'' \emph{IEEE J. Selected Areas Commun.}, vol.~35, no.~10, pp.
  2370--2382, Oct. 2017.

\bibitem{zhou2018hardware}
Z.~Zhou, N.~Ge, Z.~Wang, and S.~Chen, ``{Hardware-Efficient Hybrid Precoding
  for Millimeter Wave Systems With Multi-Feed Reflectarrays},'' \emph{IEEE
  Access}, vol.~6, pp. 6795--6806, 2018.

\bibitem{masouros2013large}
C.~Masouros, M.~Sellathurai, and T.~Ratnarajah, ``{Large-Scale MIMO
  Transmitters in Fixed Physical Spaces: The Effect of Transmit Correlation and
  Mutual Coupling},'' \emph{IEEE Trans. Commun.}, vol.~61, no.~7, pp.
  2794--2804, Jul. 2013.

\bibitem{federal2016use}
F.~C. Commission, ``{Use of Spectrum Bands Above 24 GHz For Mobile Radio
  Services},'' \emph{Fed. Regist.}, vol.~81, no. 164, pp. 58\,270--58\,308,
  2016.

\bibitem{akdeniz2014millimeter}
M.~R. Akdeniz, Y.~Liu, M.~K. Samimi, S.~Sun, S.~Rangan, T.~S. Rappaport, and
  E.~Erkip, ``{Millimeter Wave Channel Modeling and Cellular Capacity
  Evaluation},'' \emph{IEEE J. Select. Areas Commun.}, vol.~32, no.~6, pp.
  1164--1179, Jun. 2014.

\bibitem{arrebola2008multifed}
M.~Arrebola, J.~A. Encinar, and M.~Barba, ``{Multifed Printed Reflectarray with
  Three Simultaneous Shaped Beams for LMDS Central Station Antenna},''
  \emph{IEEE Trans. Antennas Propag.}, vol.~56, no.~6, pp. 1518--1527, Jun.
  2008.

\bibitem{pozar1997design}
D.~M. Pozar, S.~D. Targonski, and H.~Syrigos, ``{Design of Millimeter Wave
  Microstrip Reflectarrays},'' \emph{IEEE Trans. Antennas Propag.}, vol.~45,
  no.~2, pp. 287--296, Feb. 1997.

\bibitem{fischer2007next}
G.~Fischer, ``{Next-Generation base Station Radio Frequency Architecture},''
  \emph{Bell Labs Tech. J.}, vol.~12, no.~2, pp. 3--18, Summer 2007.

\bibitem{lin2016energy}
C.~Lin and G.~Y. Li, ``{Energy-Efficient Design of Indoor mmWave and Sub-THz
  Systems with Antenna Arrays},'' \emph{IEEE Trans. Wireless Commun.}, vol.~15,
  no.~7, pp. 4660--4672, Jul. 2016.

\bibitem{ribeiro2018energy}
L.~N. Ribeiro, S.~Schwarz, M.~Rupp, and A.~L. de~Almeida, ``{Energy Efficiency
  of mmWave Massive MIMO Precoding with Low-Resolution DACs},'' \emph{IEEE J.
  Sel. Topics Signal Process.}, vol.~12, no.~2, pp. 298--312, May 2018.

\bibitem{Rodriguez2016_MIMO_Loss}
A.~Garcia-Rodriguez, V.~Venkateswaran, P.~Rulikowski, and C.~Masouros,
  ``{Hybrid Analog-Digital Precoding Revisited Under Realistic RF Modeling},''
  \emph{IEEE Wireless Commun. Lett.}, vol.~5, no.~5, pp. 528--531, Oct. 2016.

\bibitem{hum2013reconfigurable}
S.~V. Hum and J.~Perruisseau-Carrier, ``{Reconfigurable Reflectarrays and Array
  Lenses for Dynamic Antenna Beam Control: A Review},'' \emph{IEEE Trans.
  Antennas and Propag.}, vol.~62, no.~1, pp. 183--198, Jan. 2013.

\bibitem{maloratsky2010electrically}
L.~G. Maloratsky, ``{Electrically Tunable Switched-Line Diode Phase
  Shifters},'' \emph{High Frequency Electronics}, pp. 16--21, 2010.

\bibitem{microwaves101}
\BIBentryALTinterwordspacing
``{Switched Line Phase Shifters},'' \emph{Microwave Encyclopedia}, 2020.
  [Online]. Available:
  \url{https://www.microwaves101.com/encyclopedias/switched-line-phase-shifters}
\BIBentrySTDinterwordspacing

\bibitem{alkhateeb2016frequency}
A.~Alkhateeb and R.~W. Heath, ``{Frequency Selective Hybrid Precoding for
  Limited Feedback Millimeter Wave Systems},'' \emph{IEEE Trans. Commun.},
  vol.~64, no.~5, pp. 1801--1818, May 2016.

\bibitem{sedaghat2016load}
M.~A. Sedaghat, V.~I. Barousis, R.~R. M{\"u}ller, and C.~B. Papadias, ``{Load
  Modulated Arrays: A Low-Complexity Antenna},'' \emph{IEEE Commun. Mag.},
  vol.~54, no.~3, pp. 46--52, Mar. 2016.

\bibitem{khandani2013media}
A.~K. Khandani, ``{Media-Based Modulation: A New Approach to Wireless
  Transmission},'' in \emph{ISIT}, Jul. 2013, pp. 3050--3054.

\bibitem{zang2019optimal}
G.~Zang, Y.~Cui, H.~V. Cheng, F.~Yang, L.~Ding, and H.~Liu, ``{Optimal Hybrid
  Beamforming for Multiuser Massive MIMO Systems with Individual SINR
  Constraints},'' \emph{IEEE Wireless Commun. Lett.}, Apr. 2019.

\bibitem{yu2016alternating}
X.~Yu, J.-C. Shen, J.~Zhang, and K.~B. Letaief, ``{Alternating Minimization
  Algorithms for Hybrid Precoding in Millimeter Wave MIMO Systems},''
  \emph{IEEE J. Sel. Topics Signal Process.}, vol.~10, no.~3, pp. 485--500,
  Apr. 2016.

\bibitem{castanheira2017hybrid}
D.~Castanheira, P.~Lopes, A.~Silva, and A.~Gameiro, ``{Hybrid Beamforming
  Designs for Massive MIMO Millimeter-Wave Heterogeneous Systems},'' \emph{IEEE
  Access}, vol.~5, pp. 21\,806--21\,817, Oct. 2017.

\bibitem{najafi2019cran}
M.~Najafi, V.~Jamali, D.~W.~K. Ng, and R.~Schober, ``{C-RAN with Hybrid RF/FSO
  Fronthaul Links: Joint Optimization of Fronthaul Compression and RF Time
  Allocation},'' vol.~67, no.~12, pp. 8678--8695, Dec. 2019.

\bibitem{golub2012matrix}
G.~H. Golub and C.~F. Van~Loan, \emph{{Matrix Computations}}.\hskip 1em plus
  0.5em minus 0.4em\relax JHU Press, 2012, vol.~3.

\bibitem{lau2012reconfigurable}
J.~Y. Lau, ``{Reconfigurable Transmitarray Antennas},'' Ph.D. dissertation,
  University of Toronto, 2012.

\bibitem{Pozar2009microwave}
D.~M. Pozar, \emph{{Microwave Engineering}}.\hskip 1em plus 0.5em minus
  0.4em\relax John Wiley \& Sons, 2009.

\bibitem{lee2016channel}
J.~Lee, G.-T. Gil, and Y.~H. Lee, ``{Channel Estimation via Orthogonal Matching
  Pursuit for Hybrid MIMO Systems in Millimeter Wave Communications},''
  \emph{IEEE Trans. Commun.}, vol.~64, no.~6, pp. 2370--2386, Jun. 2016.

\bibitem{zhao2017multi}
L.~Zhao, D.~W.~K. Ng, and J.~Yuan, ``{Multi-user Precoding and Channel
  Estimation for Hybrid Millimeter Wave Systems},'' \emph{IEEE J. Select. Areas
  Commun.}, vol.~35, no.~7, pp. 1576--1590, Jun. 2017.

\bibitem{mirza2018joint}
J.~Mirza, G.~Zheng, K.-K. Wong, and S.~Saleem, ``{Joint Beamforming and Power
  Optimization for D2D Underlaying Cellular Networks},'' \emph{IEEE Trans. Veh.
  Technol.}, vol.~67, no.~9, pp. 8324--8335, Sep. 2018.

\bibitem{zeng2019energy}
M.~Zeng, W.~Hao, O.~A. Dobre, and H.~V. Poor, ``{Energy-Efficient Power
  Allocation in Uplink mmwave Massive MIMO with NOMA},'' \emph{IEEE Trans. Veh.
  Technol.}, vol.~68, no.~3, pp. 3000--3004, Mar. 2019.

\bibitem{alkhateeb2014channel}
A.~Alkhateeb, O.~El~Ayach, G.~Leus, and R.~W. Heath, ``{Channel Estimation and
  Hybrid Precoding for Millimeter Wave Cellular Systems},'' \emph{IEEE J.
  Select. Topics Sig. Process.}, vol.~8, no.~5, pp. 831--846, Oct. 2014.

\bibitem{roth2018comparison}
K.~Roth, H.~Pirzadeh, A.~L. Swindlehurst, and J.~A. Nossek, ``{A Comparison of
  Hybrid Beamforming and Digital Beamforming with Low-Resolution ADCs for
  Multiple Users and Imperfect CSI},'' \emph{IEEE J. Select. Topics Sig.
  Process.}, vol.~12, no.~3, pp. 484--498, Jun. 2018.

\bibitem{bodewig2014matrix}
E.~Bodewig, \emph{{Matrix Calculus}}.\hskip 1em plus 0.5em minus 0.4em\relax
  Elsevier, 2014.

\end{thebibliography}

\end{document}